\documentclass{aa}

\usepackage{graphicx}
\usepackage{subfigure} 
\usepackage[dvipsnames]{xcolor} 
\usepackage{soul} 
\usepackage{txfonts}

\begin{document}

   \title{On the robustness of triaxial Schwarzschild modelling}
   \titlerunning{Triaxial Schwarzschild models are robust against incorrect orbit mirroring}
   \subtitle{The effects of correcting the orbit mirroring}

   \author{Sabine Thater
          \inst{1}
          \and
          Prashin Jethwa\inst{1} \and  Behzad Tahmasebzadeh\inst{2,3} \and Ling Zhu\inst{2} \and Mark den Brok\inst{4} \and Giulia Santucci\inst{5,6}  \and Yuchen Ding\inst{2,3} \and Adriano Poci\inst{7} \and  Edward Lilley\inst{1} \and P. Tim de Zeeuw\inst{8,9}  \and Alice Zocchi\inst{1}  \and  Thomas I. Maindl\inst{1,10} \and Fabio Rigamonti\inst{11,12,13} \and Glenn van de Ven\inst{1} \and Meng Yang\inst{2} \and Katja Fahrion\inst{14} 
          }

   \institute{Department of Astrophysics, University of Vienna,
              T\"urkenschanzstraße 17, 1180 Vienna\\
              \email{sabine.thater@univie.ac.at}
         \and
             Shanghai Astronomical Observatory, Chinese Academy of Sciences, 80 Nandan Road, Shanghai 200030, China
         \and   
             Department of Astronomy and Space Sciences, University of Chinese Academy of Sciences, 19A Yuquan Road, Beijing 100049, China
         \and  
             Leibniz-Institute for Astrophysics Potsdam (AIP), An der Sternwarte 16, 14482 Potsdam, Germany
        \and
        	School of Physics, University of New South Wales, NSW 2052, Australia
        \and
	        ARC center of Excellence for All Sky Astrophysics in 3 Dimensions (ASTRO 3D)
	     \and
	        Center for Extragalactic Astronomy, University of Durham, Stockton Road, Durham DH1 3LE, United Kingdom
	      \and
	        Sterrewacht Leiden, Leiden University, Postbus 9513, 2300 RA Leiden, The Netherlands
	     \and   
	        Max  Planck Institute  for  extraterrestrial Physics, Giessenbachstraße 1, 85748 Garching, Germany
	     \and
	     	SDB Science-driven Business Ltd, 85 Faneromenis Avenue, Ria Court 46, Suite 301, 6025 Larnaca, Cyprus
	     \and
	        DiSAT, Universit\`a degli Studi dell’Insubria, via Valleggio 11, 22100 Como, Italy
	     \and
	        INAF, Osservatorio Astronomico di Brera, Via E. Bianchi 46, I-23807 Merate, Italy
	     \and 
	        INFN, Sezione di Milano-Bicocca, Piazza della Scienza 3, I-20126 Milano, Italy
	      \and  
	        European Space Agency, European Space Research and Technology Centre, Keplerlaan 1, 2200 AG Noordwijk, Netherlands
             }

   \date{Received May 2nd; accepted }

  \abstract{In the past 15 years, the triaxial Schwarzschild orbit-superposition code by \citet{vandenBosch2008} has been widely applied to study the dynamics of galaxies.
  Recently, \citet{quenneville2022} reported a bug in the orbit calculation of this code, specifically in the mirroring procedure that is used to speed up the computation. We have fixed the incorrect mirroring in DYNAMITE, which is the publicly-released successor of the triaxial Schwarzschild code by \citet{vandenBosch2008}. In this study, we provide a thorough quantification of how this bug has affected the results of dynamical analyses performed with this code. We compare results obtained with the original and corrected versions of DYNAMITE, and discuss the differences in the phase-space distribution of a single orbit and in the global stellar orbit distribution, in the mass estimate of the central black hole in the highly triaxial galaxy PGC 46832, and in the measurement of intrinsic shape and enclosed mass for more than 50 galaxies. Focusing on the typical scientific applications of a Schwarzschild triaxial code, in all our tests we find that differences are negligible with respect to the statistical and systematic uncertainties. We conclude that previous results with the \citet{vandenBosch2008} triaxial Schwarzschild code are not significantly affected by the incorrect mirroring.}


   \maketitle
%

\section{Introduction}

\begin{table*}[t!]
\begin{center}
\caption[Orbit mirroring scheme]{Orbit mirroring scheme. For each mirror, the velocity components are listed, and the change in sign is clearly indicated: components marked in bold are those that have been corrected with the bugfix described in this paper. This table is adapted from \citet{quenneville2022}. Note that intermediate-axis tube orbits are not included, as they are not present in realistic models.}
\label{tab:orbits}
\begin{tabular}{ccccc}
\hline\hline
\# & octant & short-axis tube  
& long-axis tube & box \\
\hline
1 & $(x,y,z)$    & $(v_{\rm x},v_{\rm y},v_{\rm z})$ & $(v_{\rm x},v_{\rm y},v_{\rm z})$   & $(v_{\rm x},v_{\rm y},v_{\rm z})$ \\
2 & $(-x,y,z)$   & $(v_{\rm x},-v_{\rm y},\mathbf{-v_{\rm z}})$ & $(-v_{\rm x},v_{\rm y},v_{\rm z})$  & $(-v_{\rm x},v_{\rm y},v_{\rm z})$ \\
3 & $(x,-y,z)$   & $(-v_{\rm x},v_{\rm y},\mathbf{-v_{\rm z}})$ & $(\mathbf{-v_{\rm x}},v_{\rm y},-v_{\rm z})$ & $(v_{\rm x},-v_{\rm y},v_{\rm z})$ \\ 
4 & $(x,y,-z)$   & $(v_{\rm x},v_{\rm y},-v_{\rm z})$ & $(\mathbf{-v_{\rm x}},-v_{\rm y},v_{\rm z})$ & $(v_{\rm x},v_{\rm y},-v_{\rm z})$ \\ 
5 & $(-x,-y,z)$  & $(-v_{\rm x},-v_{\rm y},v_{\rm z})$ & $(\mathbf{v_{\rm x}},v_{\rm y},-v_{\rm z})$  & $(-v_{\rm x},-v_{\rm y},v_{\rm z})$ \\
6 & $(-x,y,-z)$  & $(v_{\rm x},-v_{\rm y},\mathbf{v_{\rm z}})$ & $(\mathbf{v_{\rm x}},-v_{\rm y},v_{\rm z})$  & $(-v_{\rm x},v_{\rm y},-v_{\rm z})$ \\ 
7 & $(x,-y,-z)$  & $(-v_{\rm x},v_{\rm y},\mathbf{v_{\rm z}})$ & $(v_{\rm x},-v_{\rm y},-v_{\rm z})$ & $(v_{\rm x},-v_{\rm y},-v_{\rm z})$ \\
8 & $(-x,-y,-z)$ & $(-v_{\rm x},-v_{\rm y},-v_{\rm z})$ & $(-v_{\rm x},-v_{\rm y},-v_{\rm z})$ & $(-v_{\rm x},-v_{\rm y},-v_{\rm z})$ \\ 
\hline
\end{tabular}
\end{center}
\end{table*}

Dynamical modelling is a powerful technique to study the evolution of galaxies. Traditionally this approach was restricted to spherical and axisymmetric models with analytic distribution functions depending on the integrals of motion E and L$_z$ \citep[e.g.,][]{Nagai1976,Satoh1980,Qian1995,Magorrian1998}. Observational evidence and numerical investigations showed that such models do not capture the full solution space \citep[e.g.,][]{Binney1982b}. A popular alternative is to  use the \citet{Jeans1922} equations which connect the velocity dispersions to the mass density and gravitational potential, however without the assurance that the resulting models have a non-negative distribution function (Cappellari 2008, 2020). 

\citet{Schwarzschild1979} sidestepped explicit reliance on the integrals of motion or on the Jeans equations by numerically solving the problem of populating the large variety of orbits in an assumed potential in such a way that it reproduces the mass density. As the orbits and the number of stars on it are known, the distribution of stars over position and velocity is then known everywhere in the model. The orbit occupation numbers are therefore the equivalent of the phase-space distribution function $f(x, v)\geq 0$  that is the solution to the collisionless Boltzmann equation. Elliptical galaxies can broadly be described by cored triaxial density distributions. Schwarschild (1982) demonstrated that these triaxial density distributions can be in dynamical equilibrium, even when the figure of the system is allowed to rotate slowly. \citet{Merritt1996} constructed triaxial systems with central density cusps in this way.

Since this early work, Schwarzschild’s orbit superposition method has been extended to include kinematic data \citep[e.g.,][]{Richstone1984,Rix1997,vanderMarel1998,Cretton2000}, and has been applied to spherical, axisymmetric and triaxial models with a dark halo and a central density cusp plus supermassive black hole. Rather than deprojecting the observed galaxy properties, the comparison with the numerical model is done in the observed plane, by calculating the line-of-sight integrated properties for each orbit, including, e.g., the effect of finite detector pixels, seeing convolution and internal extinction. When only the surface density is given, the solutions are non-unique \citep[e.g.,][]{Rybicki1987,Gerhard1996}. Adding kinematic constraints shrinks the range of solutions, which are computed for a range of assumed black hole masses or halo profiles. The resulting distribution of stars in phase space in many cases reveals significant structure, which reflects the different components in the observed galaxy. This decomposition is not done ad-hoc on, e.g., the surface brightness profile, but is done in phase space, constrained by all the observables, and allows exploring the formation history of the galaxy. Alternatives to Schwarzschild’s method include the made-to-measure models of \citet{Syer1996,delorenzi2008,Bovy2018}.

In the past two decades, the Schwarzschild method has been applied to multiple galaxies, initially in axisymmetric geometry and more recently also in triaxial geometry, to study their internal stellar structure \citep[e.g.,][]{Thomas2007,Cappellari2007, Ven2008, FeldmeierKrause2017, Poci2019,Jin2020,Santucci2022,Pilawa2022}, to determine their dark matter (DM) content \citep[e.g.,][]{Thomas2007,Cappellari2012,Poci2016,Santucci2022}, to weigh their central massive black holes \citep[e.g.,][]{vanderMarel1998,Verolme2002,Gebhardt2003,Valluri2004,Gebhardt2009,Krajnovic2009,Walsh2012,Rusli2013,Thater2017,Krajnovic2018,Ahn2018,Thater2019,Liepold2020,denBrok2021,Roberts2021,Thater2022,Pilawa2022} 
and to identify accreted galactic components \citep[e.g.,][]{Zhu2020,Poci2021,Zhu2022}.  
Independent implementations of the triaxial Schwarzschild method include \citet{vandenBosch2008}, \citet{Vasiliev2020} and \citet{Neureiter2021}. 
These typically use numerically computed orbit libraries on the order of $10^5$ orbits. The linear combination of these orbits is found which best represents the observed kinematics and luminosity of the galaxy under study. Each of the modelling steps requires a careful handling of state-of-the-art data and the use of high-performance computing to facilitate expensive calculations. In order to minimize the computational costs, the Schwarzschild code exploits the symmetries of the assumed galaxy potential. In a stationary, non-rotating, triaxial potential, the orbital properties are computed in one octant of phase-space, and then the remaining seven octants are filled by mirroring the orbits.

Recently, \citet{quenneville2022} pointed out that the triaxial Schwarzschild code released by \citet{vandenBosch2008} contained a bug whereby some orbit types were mirrored incorrectly. They derived the correct mirroring scheme, and tested the effect of this correction on models of the galaxy NGC 1453, finding significant changes to the $\chi^2$ distribution and best-fit parameters. This raised some concern about previously published results that were obtained with the code by \citet{vandenBosch2008}. In this work, we quantify the effect of the bug on a variety of scientific applications. We formed a collaborative team in which many members ran independent tests to obtain an unbiased, qualitative analysis.

The corrected code is available as DYNAMITE\footnote{\url{https://github.com/dynamics-of-stellar-systems/dynamite}} \citep{Jethwa2020}. This is the publicly released successor to the \citet{vandenBosch2008} code. The DYNAMITE framework provides a modern Python library with well-documented application programming interface and new functionality on top of the Schwarzschild code by \citet{vandenBosch2008}. For example, DYNAMITE can accept kinematic inputs from the BAYES-LOSVD software \citep{Falcon-Barroso2020}, which models the line-of-sight velocity distribution using a flexible, \emph{histogrammed} description; this approach may be more suitable for complex distributions than the widely-used Gauss-Hermite expansions. DYNAMITE also supports various options for solving for orbit-weight solving, including the quadratic-programming approach extolled in \citet{Vasiliev2020}.

\begin{figure*}%
\centering
\includegraphics[width=\textwidth]{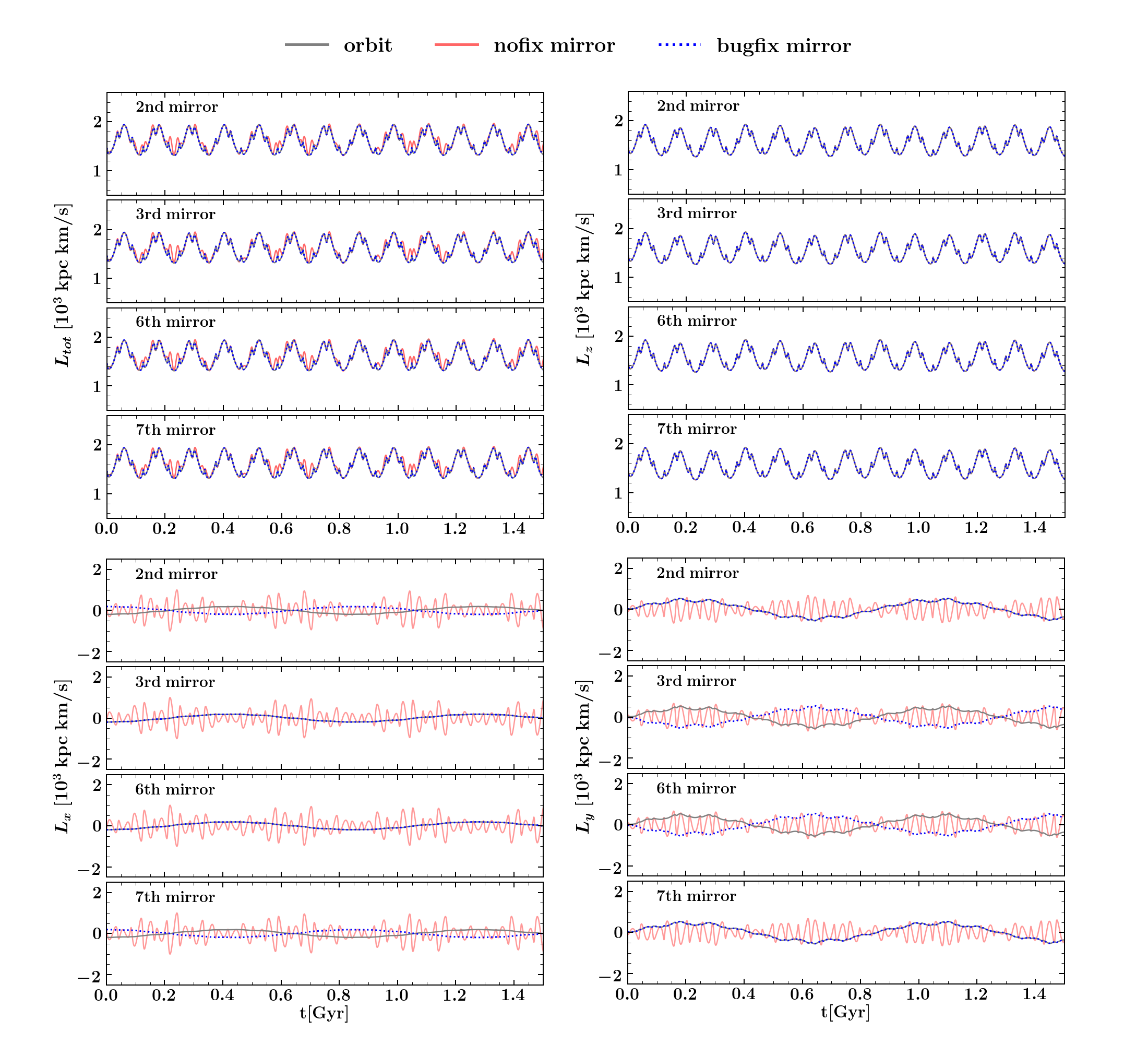} \\
\caption[]{Impact of the incorrect orbit mirroring for a single orbit. We show a typical short-axis tube orbit derived in the triaxial potential by \citet{denBrok2021}. The four panels show the evolution of $\mathrm{L_{tot}}$ (top left), $\mathrm{L_{z}}$ (top right), $\mathrm{L_{x}}$ (bottom left) and $\mathrm{L_{y}}$ (bottom right). The original orbit, correct (`bug-fix') and incorrect (`nofix') mirrors are shown by gray, blue, and red colours, respectively. We only show the four mirroring cases (out of eight) that are changed compared to the original code (see Table~\ref{tab:orbits}). In one panel, rows from top to bottom represent the 2nd, 3rd, 6th, and 7th mirrors, respectively. We show the orbit for only 25 (out of 200) revolutions to avoid crowding. The time of one orbital period is $\sim 0.06$ Gyr.}%
\label{fig:orb}%
\end{figure*}

This paper is organised as follows. In Section 2 we discuss the details of the bug and its correction. In Section 3 we show the effect of the bug-fix on several physical parameters for a large galaxy sample. In Section 4 we quantify the effect of the bug-fix for a recently published black hole mass measurement in the galaxy PGC 046832 \citep{denBrok2021}.  The inferred global orbit distributions is discussed in Section 5. In Section 6 we conclude that previously reported results are not significantly affected by the mirroring bug, and we present an outlook to on-going research with dynamical modelling.

\section{A detailed description of the orbit mirroring bug}

\begin{figure*}%
\centering

\includegraphics[width=\textwidth]{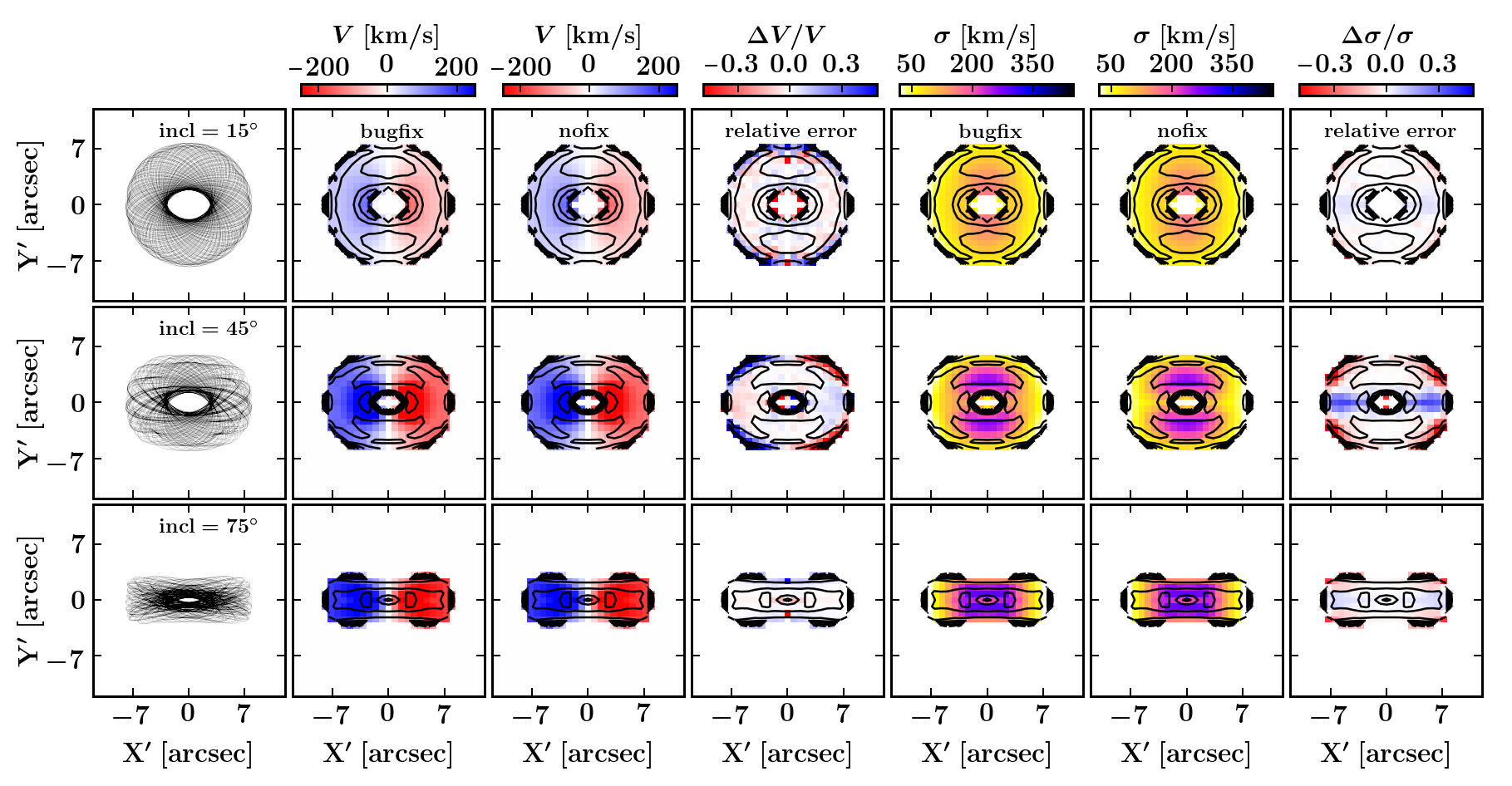}
\caption[]{Impact of the incorrect orbit mirroring for a single orbit. We show a typical short-axis tube derived in the triaxial potential by \citet{denBrok2021}. 
Columns from left to right show the orbit trajectories, the line-of-sight mean velocity $V$ with the correct (`bug-fix') and incorrect mirroring (`nofix') and the relative error between the incorrect and correct mirroring schemes $\Delta V /V$, the velocity dispersion $\sigma$ with the correct (`bug-fix') and incorrect mirroring (`nofix') and the relative error. Rows from top to bottom refer to different inclination angles of $15^{\circ}$, $45^{\circ}$ and $75^{\circ}$, respectively. The surface density of particles sampled from the orbit are overplotted as black contours. The changes in the mean velocity and velocity dispersion are up to $\sim40\%$ in some regions of the orbit space, but they are usually $\lesssim 10\%$ in regions where the star spends most of its life.}%
\label{fig:orb2}%
\end{figure*}

Stationary, non-rotating, triaxial potentials are symmetric with respect to the three principal axes. Due to this eight-fold symmetry, in the Schwarzschild code all orbital properties of regular orbits only need to be calculated in one octant and can be mirrored to the other seven. This procedure makes it possible to quickly obtain orbits starting from all octants, while only computing directly orbits that originate in one, thereby saving computational time.

Regular orbits obey a point inversion symmetry in the shape of the orbit.
Initial conditions are chosen in one octant and then integrated to obtain the orbit, and this orbit is then mirrored in the other 7 octants; if the orbit already had point symmetry then it is simply sampled more densely. However, although the spatial mirroring is always carried out identically for all orbit types, the velocity components need to be treated separately, due to the different conserved quantities associated with each orbit and the need to preserve them when the effective orientation of the orbit is flipped. Families of orbits are distinguished by which component of angular momentum, if any, they conserve. If there is no net (orbit-averaged) angular momentum conservation, or the net angular momentum is zero, then the orbit is a regular box orbit, so the signs of the mirrored velocity components just follow that of the coordinates. If the angular momentum component associated with the long axis is conserved, then it is a long-axis tube (LAT) orbit, and so in each octant the signs of $v_y$ and $v_z$ must be adjusted such that $\mathrm{L_{x}}$ remains unchanged, where
\begin{equation}
    L_x = y v_z - z v_y
\end{equation}
Similar considerations must be made for short-axis tube (SAT) orbits, which conserve $\mathrm{L_{z}}$ (sometimes a distinction is also made between inner- and outer-long-axis tube orbits, but this is not needed for the mirroring procedure). These sign changes are summarised in Table~\ref{tab:orbits}. In the incorrect mirroring scheme, 4 out of 8 velocity components of short and long-axis tube orbits had a flipped sign, meaning that the resulting mirrored orbits did not all correspond to the same (regular) orbit, and consequently should not have shared the same weight.

We analysed hundreds of individual orbits under the correct and incorrect mirroring scheme, calculated for two different galaxy potentials: a triaxial potential and a potential in the axisymmetric limit. The triaxial potential uses the luminosity model of PGC 046832 by \citet{denBrok2021} and the potential in the axisymmetric limit was derived from the luminosity model of the simulated Auriga galaxy halo 6, which was published in \citet{Zhu2020}. The second potential is close to axisymmetric as the galaxy disk dominates, but triaxiality is allowed in the model. The results of the triaxial potential are explained in the main text, while we describe the axisymmetric case in the Appendix. 

Figures \ref{fig:orb} and \ref{fig:orb2} show the impact of the incorrect orbit mirroring for a single orbit. We choose a typical short-axis tube (SAT) orbit 
in the model with triaxial potential \footnote{We show the same plots for a short-axis tube orbit in the close to axisymmetric potential in Figures \ref{fig:orbA} and \ref{fig:orb_TA}, and for a long-axis tube orbit in the same triaxial potential in Figures \ref{fig:orbB} and \ref{fig:orb_TB}}. As shown in Table~\ref{tab:orbits}, for SAT orbits, four out of eight octants had a $v_{\mathrm z}$ velocity component with an incorrect sign. 
The top left panel of Figure \ref{fig:orb} shows that the total angular momentum of these four incorrectly mirrored orbits (red line) was not equal to to that of the base orbit (grey line). After the correction (blue dotted line), this problem is solved.

Although the total angular momentum of mirrored orbits must be equal to that of the base orbit, individual components of the angular momentum may or may not change under mirroring. For a SAT orbit, which circulates around the $z$ axis, we expect all mirrored versions to have the same $\mathrm{L_{z}}$ as the base orbit. The top right panel of Figure \ref{fig:orb} shows that this is true both for the original \emph{and} corrected codes. This is in accordance with Table~\ref{tab:orbits}: since only $v_z$ was incorrect for SAT orbits, the $z$ component of the angular momentum stays unchanged by the bugfix. The other two components \emph{can} change under mirroring, but the changes must be consistent with unchanged $\mathrm{L_{z}}$ and $\mathrm{L_{tot}}$. This varies by octant: for the $2^\mathrm{nd}$ mirror, for example (top row of all panels), $\mathrm{L_{x}}$ is flipped relative to the base orbit while $\mathrm{L_{y}}$ is unchanged. In order to preserve $\mathrm{L_{tot}}$ the mirroring should only ever induce sign-changes in  components of the angular momentum. This is because, while $\mathrm{L_{tot}}$ is invariant with respect to sign-flips in the individual angular momentum components $\mathrm{L}_i$, it is \emph{not} invariant with respect to arbitrary sign-flips in the individual velocity components $v_i$. This can be seen from the formula
\begin{equation}
L_\mathrm{tot}^2 = \|\vec{x}\|^2 \|\vec{v}\|^2 - (\vec{x}\cdot\vec{v})^2,
\end{equation}
which implies that $\mathrm{L_{tot}}$ is preserved only if the combinations $(x\mathrm{L_{x}},y\mathrm{L_{y}},z\mathrm{L_{z}})$ each have the same overall sign-flip.
The short-period oscillations in the `nofix' versions of $\mathrm{L_{x}}$ and $\mathrm{L_{y}}$ are consequences of the incorrect mirroring, the exact cause of which is unknown but which is probably due to $\mathrm{L_{x}}$ and $\mathrm{L_{y}}$ no longer being constrained by the overall conservation of $\mathrm{L_{tot}}$.

In Fig. \ref{fig:orb2}, we show the trajectories of the same SAT orbit shown in Fig. \ref{fig:orb}. The trajectories were integrated for 200 times the orbital period, and we sample 50000 particles from it with equal time steps. We project the orbit with inclination angles of $15^{\circ} $ (near face-on), $45^{\circ} $ and $75^{\circ} $ (near edge-on). The projected mean velocity and velocity dispersion maps of the orbit when considering the correct (`bugfix') and incorrect (`nofix') mirroring are shown for comparison in Fig. \ref{fig:orb2}; the spatial distribution of the orbit does not depend on the mirroring of the velocity components, thus we only show it once.

Fig. \ref{fig:orb2} shows that (for this orbit) the relative error in mean velocity and velocity dispersion can reach $\sim 40\%$ in some regions, consistent with the findings of \citet{quenneville2022}. Note however that the regions with large difference are those with very low surface density of the particle. \emph{The changes are usually small ($\lesssim 10\%$) in regions of the orbit space where the star spends most of its life, and this is true for all projections.} Comparing different projections, we see that the largest errors are incurred for an intermediate inclination of $45^{\circ}$. This is somewhat surprising for this SAT orbit where only $v_{\rm z}$ has changed: the naive expectation would be that the face-on view shows the largest error, decreasing monotonically to the edge-on view. This naive logic fails to account for the fact that $v_{\rm z}$ flips sign frequently throughout this orbit. For the face-on case, the incorrect sign introduced by the  bug is on average cancelled out by the repeated sign-flips of the orbit. For the \textit{exactly} face-on and edge-on cases, we have confirmed that this orbit shows no change in its kinematic-maps after the bug-fix.

Having inspected one single orbit here, it is important to remember that Schwarzschild models are typically built from thousands of orbits. We therefore investigate in the following section, whether this change of mean velocity and velocity dispersion within a single orbit also introduces significant changes in the inferred best-fit parameters of the Schwarzschild models that use several thousands of orbits.

\begin{figure*}%
\centering
\includegraphics[width=9cm]{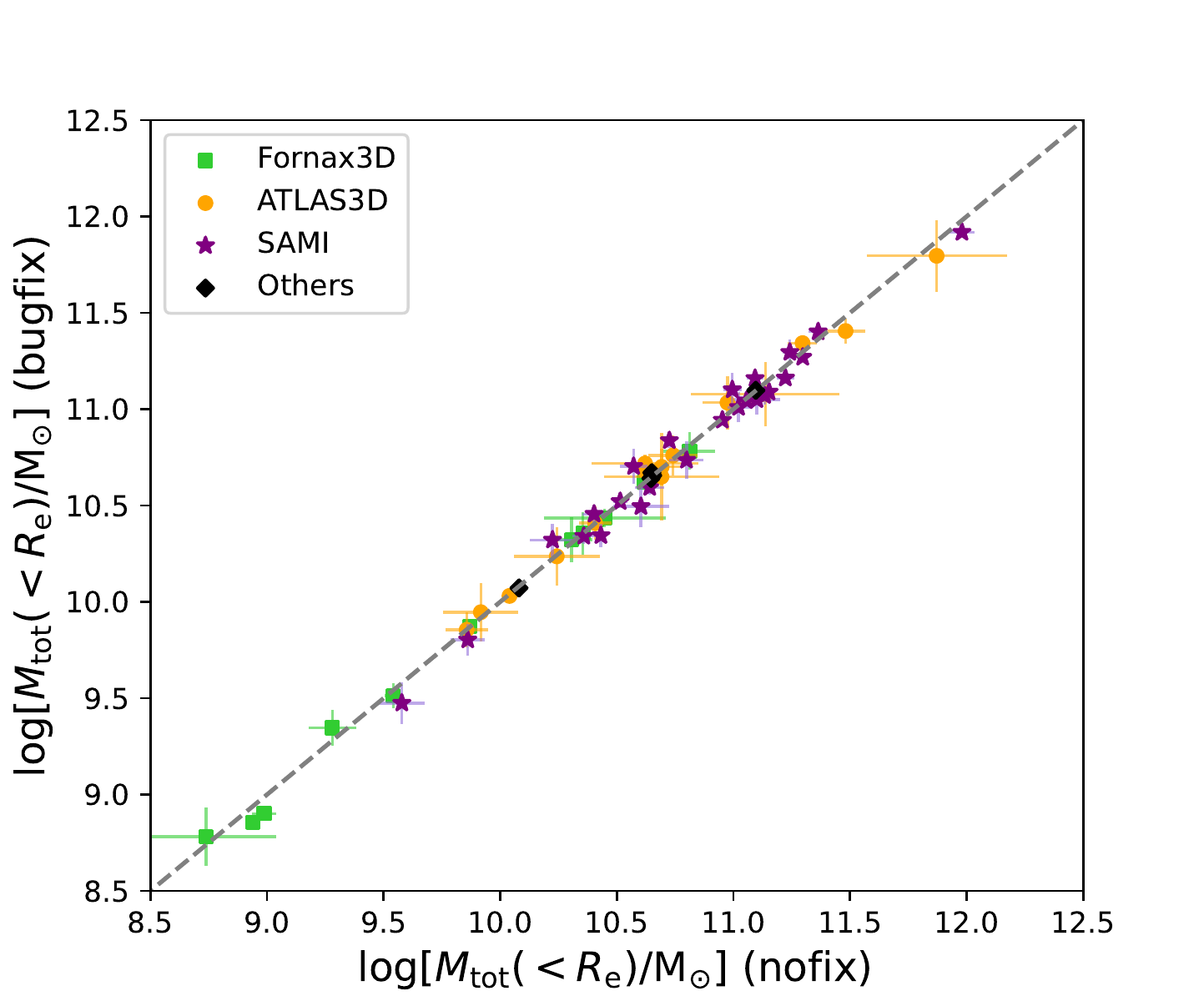}
\includegraphics[width=9cm]{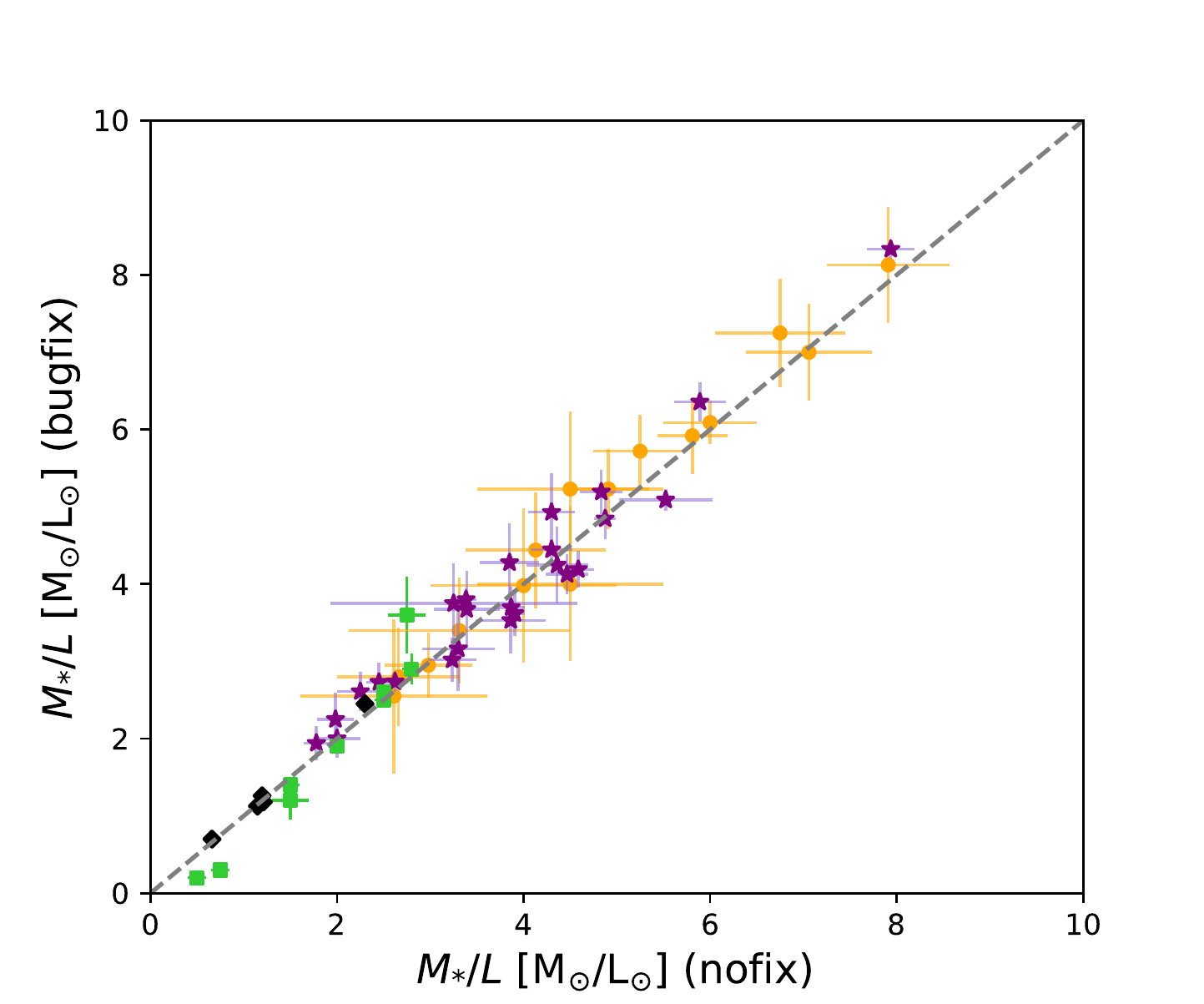}
\caption[]{Comparison of enclosed mass properties of our galaxy samples separated in total mass at one effective radius (left) and $M_*/L$ that drives the derived stellar mass (right) for correct (`bugfix') and incorrect (`nofix') orbit mirroring. The galaxy sample is a subset of Fornax3D LTGs and ETGs (green squares), SAMI passive galaxies (purple stars) and ATLAS$^{\rm 3D}$ ETGs (orange circles). Also added as black diamonds are the massive lensed ETG ESO286-G022 by \citet{Poci2022}, FCC 47 by \citet{Fahrion2019} and Thater et al. (in prep.), PGC 046832 by \citet{denBrok2021}, which is also discussed in Section 4, and the three simulated LTGs by \citet{Zhu2020}, which are discussed in Section 5. The dashed line shows the 1-1 line between the different versions. Both the derived total and stellar mass are not significantly affected by the incorrect mirroring of the orbits. Differences between the versions are within the reported statistical uncertainties of the dynamical modelling. Uncertainties were calculated by inlcuding all models within $\sqrt{2N_{\rm kin}}$ of the best-fit model.
}
\label{fig:sample1}%
\end{figure*}

\begin{figure*}%
\centering
\includegraphics[width=8.6cm]{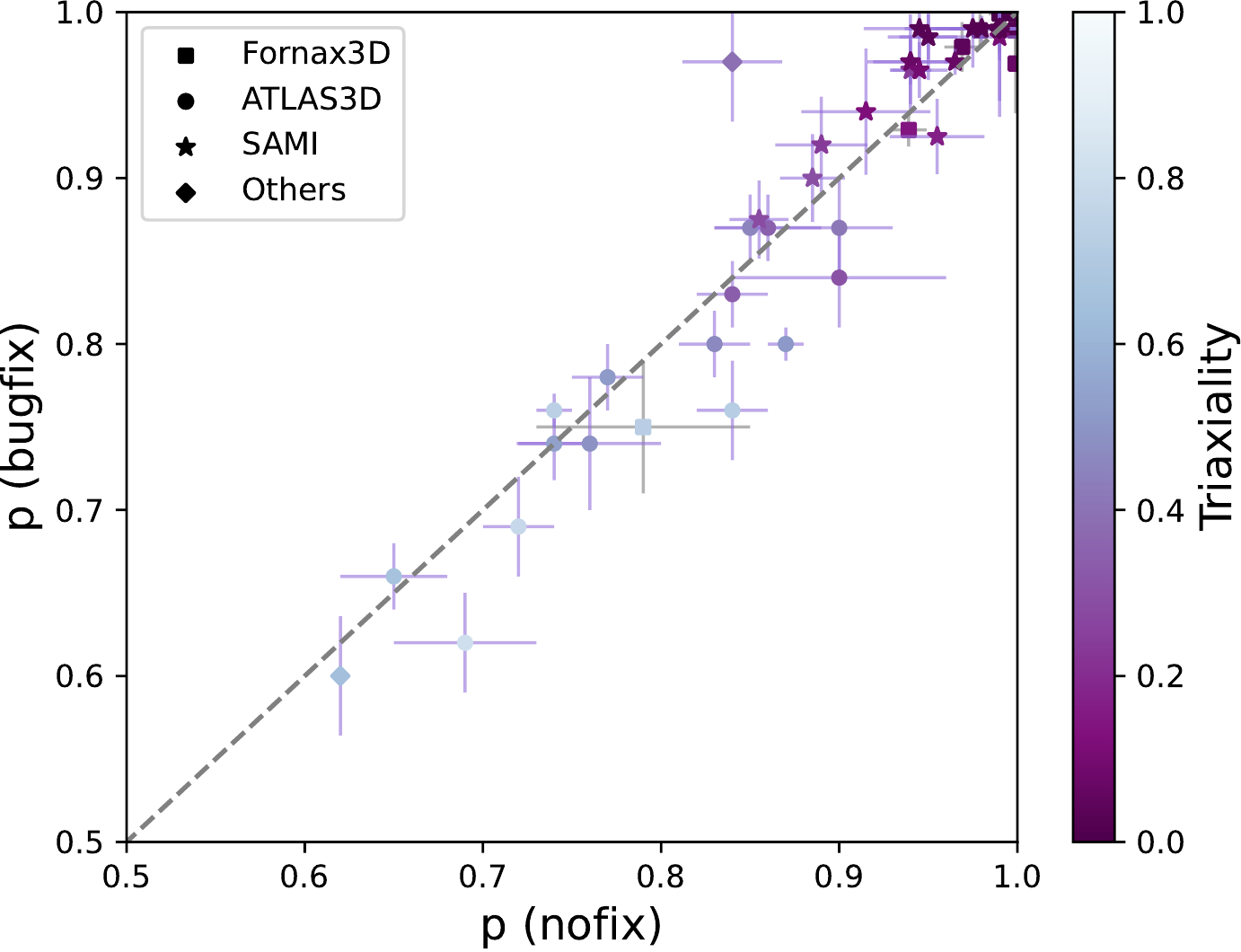}\hspace{0.4cm}
\includegraphics[width=8.6cm]{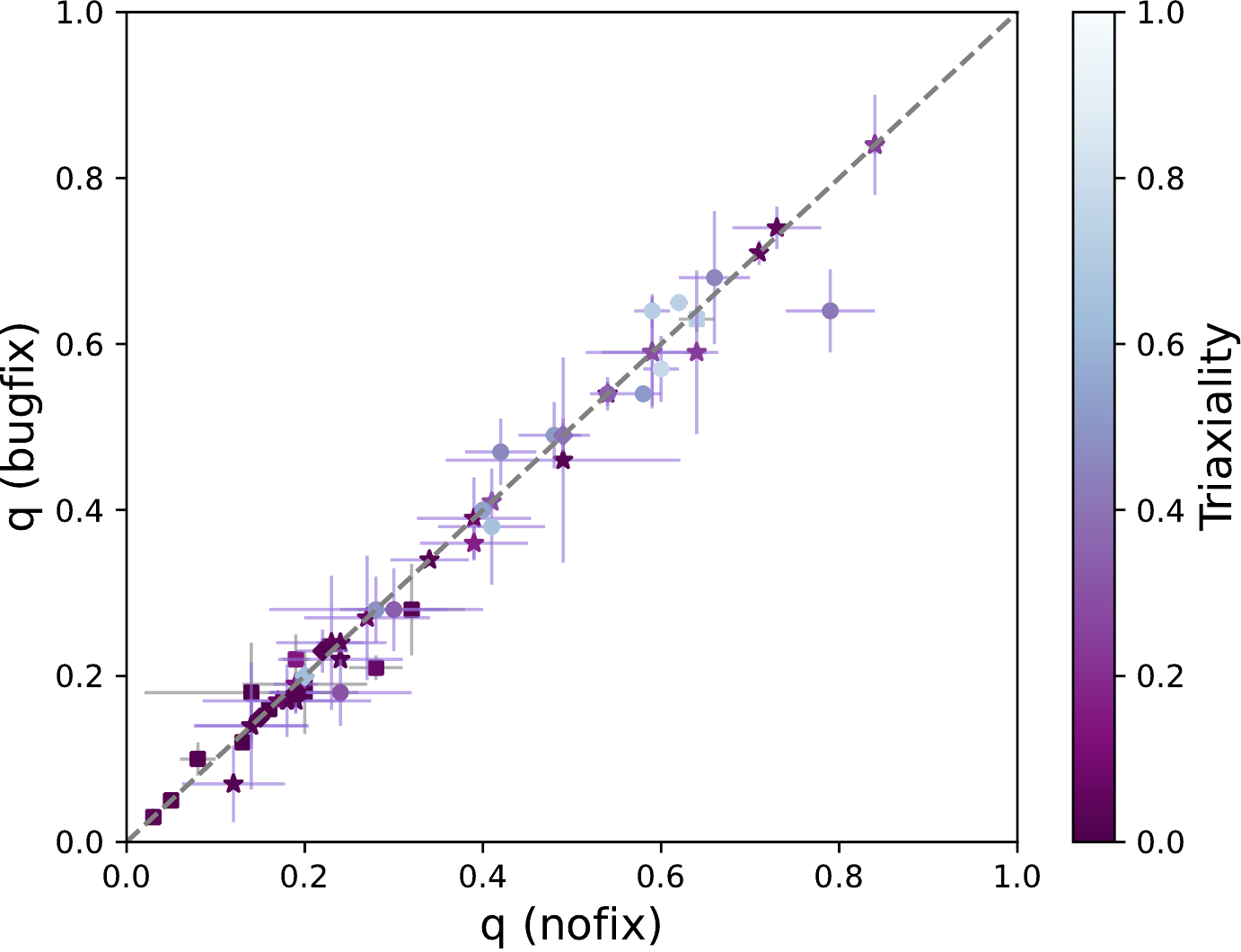}
\caption[]{Comparison of derived galaxy shape parameters for our modelled galaxies. The galaxy sample is divided into Fornax3D LTGs and ETGs (squares), SAMI passive galaxies (stars) and ATLAS$^{\rm 3D}$ ETGs (circles). Also added as diamonds are the massive lensed ETG ESO286-G022 by \citet{Poci2022}, FCC 47 by \citet{Fahrion2019} and Thater et al. (in prep.), PGC 046832 by \citet{denBrok2021}, which is also discussed in Section 4, and the three simulated LTGs by \citet{Zhu2020}, which are discussed in Section 5. Both the intrinsic intermediate-to-major axis ratio $p$ and minor-to-major axis ratio $q$ show no significant change for the majority of the galaxies. There is no clear trend of increasing discrepancy with galaxy morphology or triaxiality $T \equiv (1-p^2)/(1-q^2)$. Uncertainties were calculated by inlcuding all models within $\sqrt{2N_{\rm kin}}$ of the best-fit model

}%
	\label{fig:sample2}%
\end{figure*} 


\section{Enclosed mass and shape parameters}
Although triaxial Schwarzschild modelling is computational expensive, it is possible to study the dynamics of large galaxy samples \citep[][]{Zhu2018a,Jin2020,Santucci2022}. In order to investigate potential systematic biases due to the incorrect mirroring, we have collected data from several galaxy surveys and run the triaxial Schwarzschild code with the correct and incorrect mirroring scheme. While the previous studies had heterogeneous model setups (i.e. different numbers of orbits, different MGE assumptions), we ensured that each individual galaxy was run in the same setup for the correct and incorrect mirroring. Our analysis shows no significant discrepancies in the inferred enclosed mass and in the shape parameters of the studied galaxies. Furthermore, differences in the stellar orbit distribution caused by the incorrect mirroring are negligible compared to other systematics.

Unless otherwise stated, all models in this section are run with six free parameters: the stellar mass-to-light ratio $M_{*}/L$, intrinsic stellar axis-length ratios $p$ (long-to-short) and $q$ (intermediate-to-short), the stellar projected-to-intrinsic scale-length ratio $u$, dark matter virial mass $M_{200}$, and the dark matter concentration $c$.

\subsection{The sample}
Our studied galaxy sample is a combination of 25 passive galaxies from the SAMI Data release 3 \citep[][]{Croom2021}, 15 early-type galaxies  (ETGs) from the ATLAS$^{\rm 3D}$ galaxy survey \citep[][]{Cappellari2011} and 12 early and late-type galaxies (LTGs) from the Fornax3D survey \citep[][]{Sarzi2018}. We added 6 additional galaxies from miscellaneous other works: FCC 47 \citep[][Thater et al. in prep.]{Fahrion2019}, ESO286-G022 (\citealt{Poci2022}), PGC 046832 \citep[][]{denBrok2021} and Auriga galaxy halo 6 at inclinations 40$^{\circ}$,60$^{\circ}$ and 80$^{\circ}$ \citep[][]{Zhu2020}.

Our SAMI galaxies are a subset of the sample by \citet[][]{Santucci2022} who used the \citet{vandenBosch2008} triaxial Schwarzschild code to study the inner orbital structure and mass distribution of 161 passive galaxies. We randomly chose a subset of 25 galaxies from these galaxies with kinematic data S/N>15 and best-fit model reduced $\chi^2$<3.0. These criteria are used to avoid getting strong parameter biases due to galaxies with weak kinematic constraints or poor fit quality. We then used the same setup for the dynamical models as in \citet[][]{Santucci2022} but with correct and incorrect mirroring.

The subset of the ATLAS$^{\rm 3D}$ galaxy sample considered here was modelled with the goal to study the inner orbital structure in massive galaxies (Thater et al. in prep.) and the nature of counter-rotating galaxies (Jethwa et al. in prep.). We obtained the photometric \citep{Scott2013} and kinematic (up to h$_4$) \citep{Cappellari2011} measurements directly from the ATLAS$^{\rm 3D}$ webpage\footnote{https://www-astro.physics.ox.ac.uk/atlas3d}. The Schwarzschild models were run in the same setup as described in \citet[][]{Santucci2022} using an orbit library of $21 \times 10 \times 7$ for both tube and box orbits with a dithering of $5^3$. We fixed the dark matter concentration $c$ with the $M_{200}-c$ relation by \citet{Dutton2014} to get a better handle on the degeneracy between $M_{*}/L$ and dark matter. The ATLAS$^{\rm 3D}$ kinematics do not have the spatial resolution to constrain the central black hole, therefore we fixed its mass using the empirical $M_{\rm BH}-\sigma_{\rm e}$ relation by \citet{Bosch2016}. All models were run with the same setup for the correct and incorrect mirroring.

Finally, 6 ETGs and 6 LTGs come from the Fornax3D survey and were modelled to understand the effect of the cluster environment on the growth of cold disks (Ding et al. in prep.). We obtained the kinematics from \citet{Iodice2019}. The Fornax3D galaxies have very high-quality kinematic data: high spatial resolution; a high quality of $V$, $\sigma$, $h_3$ and $h_4$ and large kinematic data coverage (at least $2R_e$). The sample includes galaxies with morphology from highly disk-dominated to triaxial bulge dominated. Similar to the ATLAS$^{\rm 3D}$ galaxies, the black hole mass was fixed using the $M_{\rm BH}-\sigma_{\rm e}$ relation by \citet{Bosch2016} and the dark matter concentration was fixed via the $M_{200}-c$ relation. We sampled the orbits with $55 \times 11 \times 11 $ for both box and non-box orbits with dithering $5^3$.

Thus, we were able to directly compare the two different code versions over more than 50 galaxies of very diverse morphology, data quality and coverage, as well as Schwarzschild model complexity (orbit library size, fitted parameters).

\subsection{Comparison of correct and incorrect mirroring}
From a grid search of our dynamical models, we derive the best-fit parameters for each of our galaxies for the correct and incorrect mirroring. In Fig. \ref{fig:sample1} and \ref{fig:sample2}, we show a comparison of the enclosed mass properties and galaxy shapes, respectively. The enclosed mass within one effective radius is for each of our galaxies very close to the 1--1 line and thus extremely robust regarding the bug-fix. This robustness is independent of galaxy morphology, inclination or triaxiality. We want to stress here how remarkably consistent the enclosed mass measurements are, although we are using very different galaxy data sets. 

A similar behaviour is seen for the stellar mass-to-light ratio $M_{*}/L$, albeit with a larger scatter due to its degeneracy with dark matter. The $M_{*}/L$ derived with the incorrect mirroring scheme is within 10\% of the $M_{*}/L$ with the correct mirroring, thus consistent within the $M_{*}/L$ uncertainty. We conclude that both enclosed mass and $M_{*}/L$ are very robust towards the incorrect mirroring. There is strong evidence that previous dynamical mass measurements are not severely affected by the mirroring bug.

In Figure \ref{fig:sample2}, we show a comparison of the galaxy intrinsic shape parameters: the intrinsic intermediate-to-major axis ratio $p = b/a$, and the intrinsic minor-to-major axis ratio $q=c/a$. These parameters are much more difficult to constrain in dynamical models than the enclosed mass. Nevertheless, our comparison shows that $p$ and $q$ are robust against the incorrect mirroring. Almost all measurements are consistent within their uncertainties. Discrepancies are again independent of galaxy morphology, inclination or triaxiality parameter $T=(1-p^2)/(1-q^2)$. As $p$ and $q$ enter $T$ via their squares, the discrepancy in the derived $T$ values shows more scatter than in either individual axis-ratio, but again there is no clear trend with galaxy properties. Some galaxies are more driven to a prolate shape, others more oblate, while others did not change. From this comparison, we conclude that galaxy intrinsic shapes in previous results likely do not suffer from systematic biases. We point out that intrinsic shape parameters depend on the quality of the data, and larger discrepancies are found for kinematic data with S/N$<15$.


\section{A black hole mass estimate in a triaxial galaxy}
\label{sec:bh}

\citet{denBrok2021}

used the triaxial code of \citet{vandenBosch2008} in its original version to model the VLT/MUSE observations of PGC 046832. PGC 046832 is the brightest cluster galaxy in one of the subclusters of the Shapley Supercluster, at a distance of $\sim\!\!200$ Mpc. Because of its complex structure, this galaxy poses a challenge for modellers, requiring several inversions in the direction of its angular momentum and a radial change in triaxiality.

The dynamical modeling by \citet{denBrok2021} showed that a) the black hole mass determined with the triaxial Schwarzschild models was lower than the one determined using axisymmetric models and that b) the intrinsic shape of the galaxy changes from almost prolate in the centre to almost oblate in the outer parts. Here we show that correcting the orbit mirroring in the triaxial Schwarzschild code by \citet{vandenBosch2008} does not lead to a detection of the central supermassive black hole and only marginally changes the best-fit viewing angles. We have added the stellar orbit distribution (see also Section 5) of PGC 046832 to the Appendix (Fig.~\ref{fig:pgc_circ}). Changes in these plots compared to \citet{denBrok2021} are not significant.

\subsection{Black hole mass}
To determine the direct influence of the orbit mirroring correction on the black hole mass of PGC 046832, we assume the same viewing angles as used by \citet{denBrok2021} and the same dark matter halo mass. We re-run the Schwarzschild models in the exact same setup as in \citet{denBrok2021}, but additionally supplement the grid with points at $M/L$ between 2.9 and 3.0, as the $\chi^2$ contours imply a 4\% lower $M/L$.

\begin{figure}
  \includegraphics[angle=0, width=0.49\textwidth]{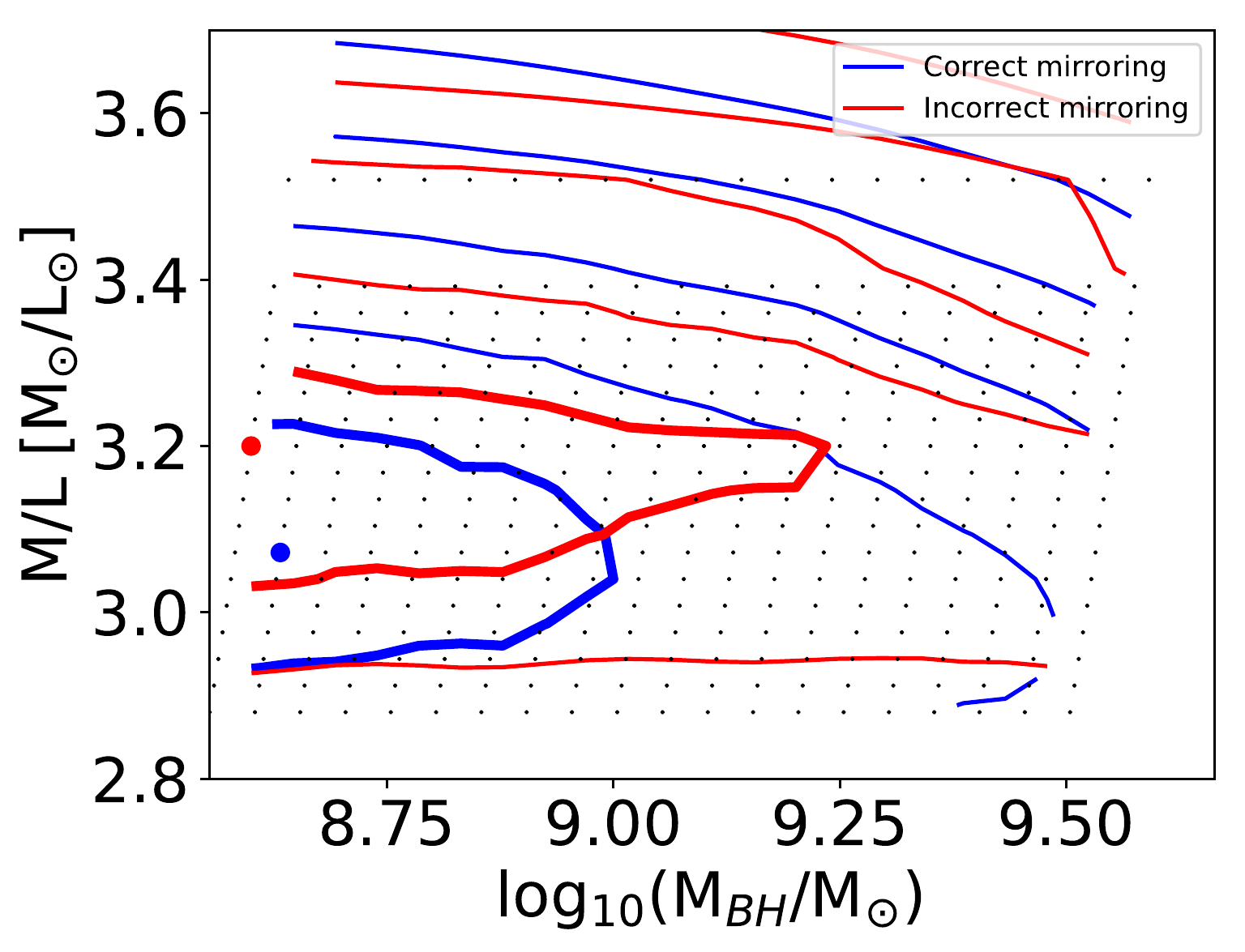}
  \caption{Contours describing the $\chi^2$ surfaces at different black hole masses $M_{\rm BH}$ and mass to light ratios $M/L$. Blue contours show the $\chi^2$ surface using the original version of the Schwarzschild code, and were previously presented in  \citet{denBrok2021}. Red contours show the $\chi^2$ surface after correcting the mirroring bug in the code. Grey dots show the locations at which models were calculated. The thick contours contain models within $\sqrt{2N_{\mathrm{kin}}}$ of the best-fit model. The contours do not close for lower-mass black holes and we thus obtain an upper limit. The sphere of influence of a $10^9\,M_{\odot}$ black hole is about 0.05 arcsec which is well below the spatial resolution of the data (seeing fwhm $\approx$ 0.67 arcsec).
  }
  \label{fig:pgc046832_contours}
  \end{figure}

We show the $\chi^2$ contours in Fig. \ref{fig:pgc046832_contours}, where the contours from the correctly mirrored code are given in red, and those from the incorrectly mirrored code in blue. The blue contours correspond to the black contours in Fig. 10 of \citet{denBrok2021}.  The correction of the orbit mirroring bug does not lead to a black hole detection. The derived $M/L$ decreases by 4\%, but this range is consistent within the $3\sigma$ confidence interval of the measurement. The red contours are also noticeably more round than the blue contours, indicating a change in the $\chi^2$ distribution. When inspecting the model kinematic maps of the best-fitting models, we did not notice any strong differences.

\subsection{Viewing angles}
\citet{denBrok2021} assumed that the density of the galaxy could be modelled as a sum of aligned concentric Gaussians with different axis ratios $p_i$ and $q_i$ and scale lengths $\sigma_i$. This assumption allows an analytic deprojection of the observed light distribution given a set of viewing angles. Viewing angles therefore affect the gravitational potential in which orbits are calculated, and thus can be dynamically constrained. \citet{denBrok2021} used Schwarzschild models to constrain the viewing angles as ($\theta, \phi, \Psi$) = ($61.0_{-1.8}^{+3.8}$,$-59.4_{-2.4}^{+4.8}$,$74.9_{-1.2}^{+1.0}$) in degrees.

We re-run Schwarzschild models with the same assumptions as in \citet{denBrok2021}, i.e. fixed black hole mass and 3 different masses for the DM halo, to recreate the grid shown in their Fig. 6. We also explored the parameter space with a free DM halo mass and black hole mass using the same Gaussian sampling approach used in that paper. For the best-fit viewing angles we find ($\theta, \phi, \Psi$) = ($60.2_{-2.5}^{+3.2}$,$-61.9_{-1.5}^{+3.9}$,$75.9_{-1.2}^{+0.1}$), consistent with those obtained with the previous version of the code.

The deprojection of the MGE using these new viewing angles does not lead to a significantly different intrinsic shape. After correcting the mirroring bug, a preference towards a massive black hole with mass $\log(M_{\rm BH}/$M$_{\odot}) \lesssim 8$ is still predicted by the fits. This does not mean that the new models shows no changes at all; the mass of the dark matter halo, expressed as a dimensionless scaling of the MGE mass, is higher than before ($2027_{-591}^{+286}$ versus $1477_{-638}^{+385} $), whereas the stellar $M/L$ is somewhat lower ($3.08_{-0.08}^{0.08}$ versus $3.18_{-0.09}^{+0.04}$). The total galaxy mass within the radii that can be constrained by the kinematics is however consistent as was also found for other galaxies in Fig.~\ref{fig:sample1}. We note that the changes on these quantities are likely caused by the intrinsic degeneracy between dark matter and stellar $M/L$, and the limited kinematic field-of-view that cannot constrain the dark matter very well.

\begin{figure*}%
\centering

\includegraphics[width=12cm]{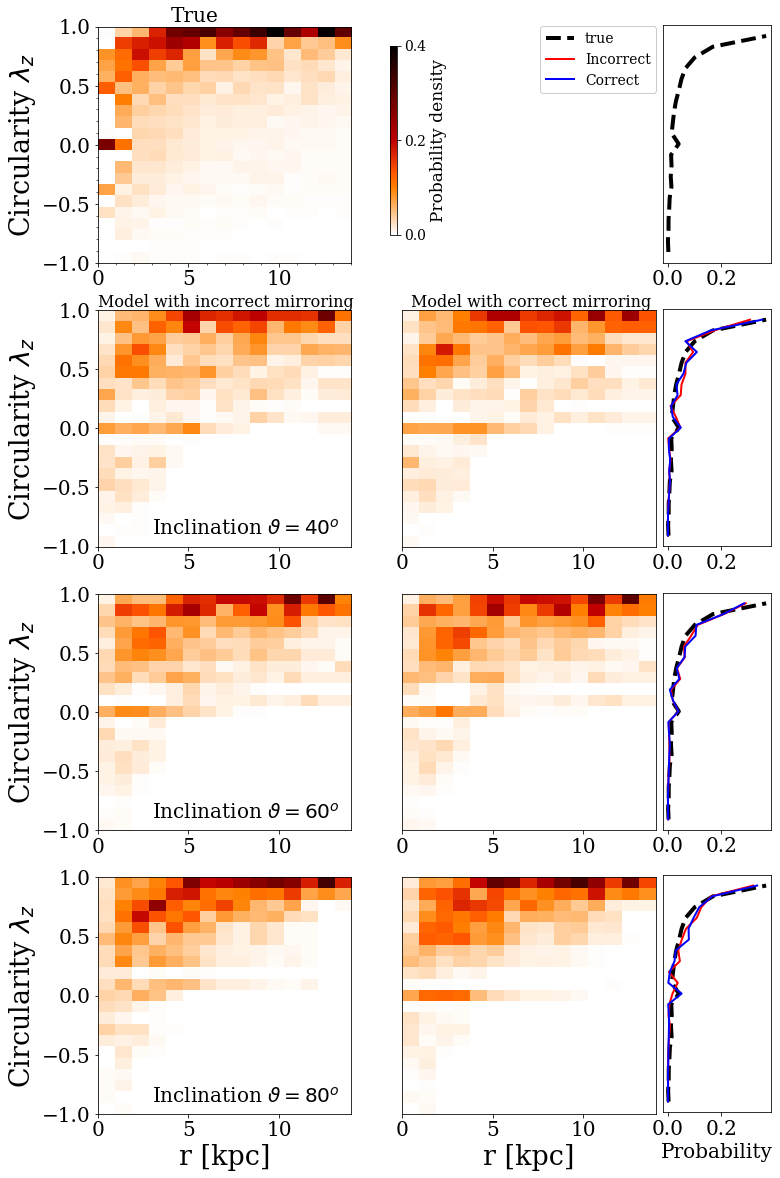}
\caption[]{The stellar orbit distribution in intrinsic radius $r$ vs. circularity $\lambda_z \equiv \overline{L_z} /(r \times \overline{V_c})$ for mock spiral galaxies with inclination angles of $40^{\circ}$, $60^{\circ}$ and $80^{\circ}$ from top to bottom. The top panel shows the true stellar orbit distribution from the simulation. In the following three rows, the left panels are for models with incorrect mirroring, the right panels are for models with correct mirroring. The side panels compare the $\lambda_z$ distribution for all the orbits at $r<15$ kpc. The orbit distributions from the models with correct and incorrect mirroring are nearly identical, and they both present some difference with respect to the true distribution. The differences caused by the incorrect mirroring are negligible compared to the other systematic errors.}	
	\label{fig:circ}%
\end{figure*}

\subsection{Other black hole mass measurements}
Few black hole mass measurements have been derived with the triaxial Schwarzschild code by \citet{vandenBosch2008}. It has been applied to the two mildly triaxial fast-rotating early-type galaxies M32 and NGC 3379 \citep{vandenBosch2010}, the moderately triaxial early-type galaxy NGC 3998 \citep{Walsh2012}, the nuclear star cluster in the Milky Way \citep{FeldmeierKrause2017} and the ultracompact dwarf galaxy M59-UCD3 \citep{Ahn2018}. Some of these studies resulted in black hole mass measurements that were inconsistent with other methods. Our investigation suggests that this inconsistency is not driven by the incorrect mirroring bug, but by other systematics, e.g. radially varying versus constant mass-to-light ratio \citep{Thater2017,Thater2019,Thater2022}, the inclusion of dark matter into the models \citep{Gebhardt2009,Rusli2013,Thater2022} or other assumptions of the modelling techniques. Furthermore, the incorrect mirroring bug is not present in the Leiden version of the axisymmetric Schwarzschild code that was used to derive several black hole mass measurements \citep[e.g.,][and the axisymmetric measurement in \citealt{denBrok2021}]{Krajnovic2009,Krajnovic2018,Thater2017,Thater2019,Thater2022} . Cross-checks between the different code versions are extremely valuable to find systematic differences and mistakes in the codes.

\section{Stellar orbit distribution}

In addition to constraints on the gravitational potential, a useful result of Schwarzschild modelling is the stellar orbit distribution. This distribution is often shown in the space of circularity, $\lambda_{\rm z} = \overline{L_z} /(r \times \overline{V_c})$, i.e. the orbit angular momentum $\overline{L_z}$ normalized by the angular momentum of a circular orbit with the same binding energy \citep{Zhu2018b}. This means that |$\lambda_{\rm z}$| = 1 represents highly-rotating short-axis tube orbits (circular orbits), while $\lambda_{\rm z} = 0$ represents mostly dynamically hot box or radial orbits. The circularity distribution has been used in the past to disentangle dynamical cold, warm and hot components and learn about the accretion history of nearby galaxies \citep[e.g.,][]{Zhu2018b,Zhu2020,Zhu2022,Fahrion2019,Poci2019}. 

We construct triaxial Schwarzschild models for three mock late-type galaxies, created from the same simulation Auriga halo 6, but with inclination angles of $40^{\circ}$, $60^{\circ}$ and $80^{\circ}$, respectively. The advantage of using mock galaxies is that we know the underlying true orbit distribution. The creation of the mock data and dynamical models were performed in the same setup described in \citet[][]{Zhu2020}, but using the triaxial Schwarzschild code version with correct and incorrect mirroring.

Figure \ref{fig:circ} illustrates the stellar orbit distributions in radius $r$ vs. circularity $\lambda_{\rm z}$ of the three mock galaxies derived by our models with the correct and incorrect mirroring. For each mock galaxy, we only use the best-fitting model. Their orbit distributions are compared to the true orbit distribution from the simulation. The darker colour indicates a higher phase space density. Being LTGs, the mock galaxies are dominated by a dynamical cold component ($\lambda_{\rm z}$ = 1) which is accompanied by a central hot component, the bulge of the galaxy. The proportion of cold and hot components is well reproduced for both the correct and incorrect mirroring. Differences caused by the incorrect mirroring are small compared to the systematic errors of the model (as inferred by departures from the ground truth), and independent of the inclination angle. The side panels in Fig. \ref{fig:circ} confirm that the trends in the orbit fractions are almost the same for the correct and incorrect mirroring.

Although the line-of-sight mean velocity ($V$) and velocity dispersion ($\sigma$) maps of each single orbit change with the incorrect mirroring in the code, we find that the orbit distributions in the $r$ vs. $\lambda_{\rm z}$ plane of the best-fitting models do not noticeably change.
This is because, as shown in Fig. 2, for each orbit larger differences in $\sigma$ occur in the areas with low density. In areas with larger surface density (where orbit contributes more light to the model)  the difference in $\sigma$ is small.  
Our model is a combination of many orbits. For orbits with certain ($V$, $\sigma$) maps an alternative combination of orbits with similar ($V$, $\sigma$) maps might be found in the library calculated with the incorrect code version, although not orbits with exactly the same combination of integrals of motion ($E$, L$_z$, $I_3$). When fitting to the kinematic maps, the two version of models may pick different orbits regarding to their $E$, L$_z$, $I_3$, but prefer orbits with similar ($V$, $\sigma$), thus similar $\lambda_{\rm z}$. 

Note that these results are not only true for the three mock galaxies. The stellar orbit distributions in the $r$ vs. $\lambda_{\rm z}$ plane for all the galaxies we have remodelled using the correct mirroring in Section 4 are very similar to those obtained using the non-corrected code.


\section{Conclusion and future work}
Recently, \citet{quenneville2022} reported a bug in the original triaxial Schwarzschild code published by \citet{vandenBosch2008} where some orbits had been incorrectly mirrored. After correcting the mirroring in our open-source triaxial Schwarzschild code (DYNAMITE, \citealt{Jethwa2020}), we carefully checked for systematic changes regarding the estimate of a black hole mass, enclosed mass, intrinsic shape and the stellar orbit distribution. We noticed small effects in the shape of the $\chi^2$ distribution, but the best-fit parameters with the correct and incorrect mirroring were in almost all cases consistent within their uncertainties. We did not see any noticeable trends with galaxy morphology, inclination angle or triaxiality. 

\citet{denBrok2021} used the original triaxial Schwarzschild code to derive the black hole mass in PGC 046832 and found significant discrepances between the results from the triaxial Schwarzschild code and axisymmetric dynamical methods. We have re-analysed the galaxy with the corrected triaxial Schwarzschild code, but obtained consistent results with \citet{denBrok2021}.

We can therefore conclude that the incorrect mirroring did not systematically bias previous results obtained with the triaxial Schwarzschild code by \citet{vandenBosch2008}. 

Several other developments are planned for DYNAMITE in the near future. We will implement the \emph{orbit-colouring} technique to incorporate stellar population information \citep{Poci2019,Zhu2020} and offer more sophisticated parameter search algorithms \citep{Gration2019}. Finally, we recognize the need for a more thorough treatment of uncertainties for orbit-based modelling. An important step in this direction has been made in \citet{Lipka2021}, who presented a novel technique to optimize the amount of regularization used for orbit-weight solving, which can have a significant impact on constraints on physical parameters such as galaxy mass. How to account for uncertatinies on the orbit-weights themselves is an open question which has been largely ignored due to the difficulty of assigning meaningful uncertainties in the high-dimensional and degenerate space of orbit-weights, however, getting a handle on these uncertainties is vital if we wish to associate clumpiness in an orbit-distribution to the presence of merged galactic components. Addressing these concerns is the focus of our future research.

\section{Acknowledgements}
We thank M. Quenneville and their group for finding and reporting the bug in the original code. Their work has motivated us to re-evaluate previous results and report our work in this collaborative study. We also thank Eugene Vasiliev, the author of the SMILE orbit-superposition code, for many fruitful and constructive discussions. This project has received funding from the European Research Council (ERC) under the European Union’s Horizon 2020 research and innovation programme under grant agreement No 724857 (Consolidator Grant ArcheoDyn). The computational results presented have been achieved (in part) using the Vienna Scientific Cluster (VSC, \url{https://vsc.ac.at}). AP is supported by the Science and Technology Facilities Council through the Durham Astronomy Consolidated Grant 2020–2023 (ST/T000244/1).

\bibliography{papers}

\begin{thebibliography}{65}
\expandafter\ifx\csname natexlab\endcsname\relax\def\natexlab#1{#1}\fi

\bibitem[{{Ahn} {et~al.}(2018){Ahn}, {Seth}, {Cappellari}, {Krajnovi{\'c}},
  {Strader}, {Voggel}, {Walsh}, {Bahramian}, {Baumgardt}, {Brodie},
  {Chilingarian}, {Chomiuk}, {den Brok}, {Frank}, {Hilker}, {McDermid},
  {Mieske}, {Neumayer}, {Nguyen}, {Pechetti}, {Romanowsky}, \&
  {Spitler}}]{Ahn2018}
{Ahn}, C.~P., {Seth}, A.~C., {Cappellari}, M., {et~al.} 2018, \apj, 858, 102

\bibitem[{{Binney}(1982)}]{Binney1982b}
{Binney}, J. 1982, \araa, 20, 399

\bibitem[{{Bovy} {et~al.}(2018){Bovy}, {Kawata}, \& {Hunt}}]{Bovy2018}
{Bovy}, J., {Kawata}, D., \& {Hunt}, J. A.~S. 2018, \mnras, 473, 2288

\bibitem[{{Cappellari} {et~al.}(2007){Cappellari}, {Emsellem}, {Bacon},
  {Bureau}, {Davies}, {de Zeeuw}, {Falc{\'o}n-Barroso}, {Krajnovi{\'c}},
  {Kuntschner}, {McDermid}, {Peletier}, {Sarzi}, {van den Bosch}, \& {van de
  Ven}}]{Cappellari2007}
{Cappellari}, M., {Emsellem}, E., {Bacon}, R., {et~al.} 2007, \mnras, 379, 418

\bibitem[{{Cappellari} {et~al.}(2011){Cappellari}, {Emsellem}, {Krajnovi{\'c}},
  {McDermid}, {Scott}, {Verdoes Kleijn}, {Young}, {Alatalo}, {Bacon}, {Blitz},
  {Bois}, {Bournaud}, {Bureau}, {Davies}, {Davis}, {de Zeeuw}, {Duc},
  {Khochfar}, {Kuntschner}, {Lablanche}, {Morganti}, {Naab}, {Oosterloo},
  {Sarzi}, {Serra}, \& {Weijmans}}]{Cappellari2011}
{Cappellari}, M., {Emsellem}, E., {Krajnovi{\'c}}, D., {et~al.} 2011, \mnras,
  413, 813

\bibitem[{{Cappellari} {et~al.}(2012){Cappellari}, {McDermid}, {Alatalo},
  {Blitz}, {Bois}, {Bournaud}, {Bureau}, {Crocker}, {Davies}, {Davis}, {de
  Zeeuw}, {Duc}, {Emsellem}, {Khochfar}, {Krajnovi{\'c}}, {Kuntschner},
  {Lablanche}, {Morganti}, {Naab}, {Oosterloo}, {Sarzi}, {Scott}, {Serra},
  {Weijmans}, \& {Young}}]{Cappellari2012}
{Cappellari}, M., {McDermid}, R.~M., {Alatalo}, K., {et~al.} 2012, \nat, 484,
  485

\bibitem[{{Cretton} {et~al.}(2000){Cretton}, {Rix}, \& {de
  Zeeuw}}]{Cretton2000}
{Cretton}, N., {Rix}, H.-W., \& {de Zeeuw}, P.~T. 2000, \apj, 536, 319

\bibitem[{{Croom} {et~al.}(2021){Croom}, {Owers}, {Scott}, {Poetrodjojo},
  {Groves}, {van de Sande}, {Barone}, {Cortese}, {D'Eugenio}, {Bland-Hawthorn},
  {Bryant}, {Oh}, {Brough}, {Agostino}, {Casura}, {Catinella}, {Colless},
  {Cecil}, {Davies}, {Drinkwater}, {Driver}, {Ferreras}, {Foster},
  {Fraser-McKelvie}, {Lawrence}, {Leslie}, {Liske}, {L{\'o}pez-S{\'a}nchez},
  {Lorente}, {McElroy}, {Medling}, {Obreschkow}, {Richards}, {Sharp}, {Sweet},
  {Taranu}, {Taylor}, {Tescari}, {Thomas}, {Tocknell}, \&
  {Vaughan}}]{Croom2021}
{Croom}, S.~M., {Owers}, M.~S., {Scott}, N., {et~al.} 2021, \mnras, 505, 991

\bibitem[{{de Lorenzi} {et~al.}(2008){de Lorenzi}, {Gerhard}, {Saglia},
  {Sambhus}, {Debattista}, {Pannella}, \& {M{\'e}ndez}}]{delorenzi2008}
{de Lorenzi}, F., {Gerhard}, O., {Saglia}, R.~P., {et~al.} 2008, \mnras, 385,
  1729

\bibitem[{{den Brok} {et~al.}(2021){den Brok}, {Krajnovi{\'c}}, {Emsellem},
  {Brinchmann}, \& {Maseda}}]{denBrok2021}
{den Brok}, M., {Krajnovi{\'c}}, D., {Emsellem}, E., {Brinchmann}, J., \&
  {Maseda}, M. 2021, \mnras, 508, 4786

\bibitem[{{Dutton} \& {Macci{\`o}}(2014)}]{Dutton2014}
{Dutton}, A.~A. \& {Macci{\`o}}, A.~V. 2014, \mnras, 441, 3359

\bibitem[{{Fahrion} {et~al.}(2019){Fahrion}, {Lyubenova}, {van de Ven},
  {Leaman}, {Hilker}, {Mart{\'\i}n-Navarro}, {Zhu}, {Alfaro-Cuello}, {Coccato},
  {Corsini}, {Falc{\'o}n-Barroso}, {Iodice}, {McDermid}, {Sarzi}, \& {de
  Zeeuw}}]{Fahrion2019}
{Fahrion}, K., {Lyubenova}, M., {van de Ven}, G., {et~al.} 2019, \aap, 628, A92

\bibitem[{{Falc{\'o}n-Barroso} \& {Martig}(2021)}]{Falcon-Barroso2020}
{Falc{\'o}n-Barroso}, J. \& {Martig}, M. 2021, \aap, 646, A31

\bibitem[{{Feldmeier-Krause} {et~al.}(2017){Feldmeier-Krause}, {Zhu},
  {Neumayer}, {van de Ven}, {de Zeeuw}, \& {Sch{\"o}del}}]{FeldmeierKrause2017}
{Feldmeier-Krause}, A., {Zhu}, L., {Neumayer}, N., {et~al.} 2017, \mnras, 466,
  4040

\bibitem[{{Gebhardt} {et~al.}(2003){Gebhardt}, {Richstone}, {Tremaine},
  {Lauer}, {Bender}, {Bower}, {Dressler}, {Faber}, {Filippenko}, {Green},
  {Grillmair}, {Ho}, {Kormendy}, {Magorrian}, \& {Pinkney}}]{Gebhardt2003}
{Gebhardt}, K., {Richstone}, D., {Tremaine}, S., {et~al.} 2003, \apj, 583, 92

\bibitem[{{Gebhardt} \& {Thomas}(2009)}]{Gebhardt2009}
{Gebhardt}, K. \& {Thomas}, J. 2009, \apj, 700, 1690

\bibitem[{{Gerhard} \& {Binney}(1996)}]{Gerhard1996}
{Gerhard}, O.~E. \& {Binney}, J.~J. 1996, \mnras, 279, 993

\bibitem[{Gration \& Wilkinson(2019)}]{Gration2019}
Gration, A. \& Wilkinson, M.~I. 2019, \mnras, 485, 4878

\bibitem[{{Iodice} {et~al.}(2019){Iodice}, {Sarzi}, {Bittner}, {Coccato},
  {Costantin}, {Corsini}, {van de Ven}, {de Zeeuw}, {Falc{\'o}n-Barroso},
  {Gadotti}, {Lyubenova}, {Mart{\'\i}n-Navarro}, {McDermid}, {Nedelchev},
  {Pinna}, {Pizzella}, {Spavone}, \& {Viaene}}]{Iodice2019}
{Iodice}, E., {Sarzi}, M., {Bittner}, A., {et~al.} 2019, \aap, 627, A136

\bibitem[{{Jeans}(1922)}]{Jeans1922}
{Jeans}, J.~H. 1922, \mnras, 82, 122

\bibitem[{{Jethwa} {et~al.}(2020){Jethwa}, {Thater}, {Maindl}, \& {Van de
  Ven}}]{Jethwa2020}
{Jethwa}, P., {Thater}, S., {Maindl}, T., \& {Van de Ven}, G. 2020,
  Astrophysical Source Code Library, record ascl:2011.007

\bibitem[{{Jin} {et~al.}(2020){Jin}, {Zhu}, {Long}, {Mao}, {Wang}, \& {van de
  Ven}}]{Jin2020}
{Jin}, Y., {Zhu}, L., {Long}, R.~J., {et~al.} 2020, \mnras, 491, 1690

\bibitem[{{Krajnovi{\'c}} {et~al.}(2018){Krajnovi{\'c}}, {Cappellari},
  {McDermid}, {Thater}, {Nyland}, {de Zeeuw}, {Falc{\'o}n-Barroso}, {Khochfar},
  {Kuntschner}, {Sarzi}, \& {Young}}]{Krajnovic2018}
{Krajnovi{\'c}}, D., {Cappellari}, M., {McDermid}, R.~M., {et~al.} 2018,
  \mnras, 477, 3030

\bibitem[{{Krajnovi{\'c}} {et~al.}(2009){Krajnovi{\'c}}, {McDermid},
  {Cappellari}, \& {Davies}}]{Krajnovic2009}
{Krajnovi{\'c}}, D., {McDermid}, R.~M., {Cappellari}, M., \& {Davies}, R.~L.
  2009, \mnras, 399, 1839

\bibitem[{{Liepold} {et~al.}(2020){Liepold}, {Quenneville}, {Ma}, {Walsh},
  {McConnell}, {Greene}, \& {Blakeslee}}]{Liepold2020}
{Liepold}, C.~M., {Quenneville}, M.~E., {Ma}, C.-P., {et~al.} 2020, \apj, 891,
  4

\bibitem[{{Lipka} \& {Thomas}(2021)}]{Lipka2021}
{Lipka}, M. \& {Thomas}, J. 2021, \mnras, 504, 4599

\bibitem[{{Magorrian} {et~al.}(1998){Magorrian}, {Tremaine}, {Richstone},
  {Bender}, {Bower}, {Dressler}, {Faber}, {Gebhardt}, {Green}, {Grillmair},
  {Kormendy}, \& {Lauer}}]{Magorrian1998}
{Magorrian}, J., {Tremaine}, S., {Richstone}, D., {et~al.} 1998, \aj, 115, 2285

\bibitem[{{Merritt} \& {Fridman}(1996)}]{Merritt1996}
{Merritt}, D. \& {Fridman}, T. 1996, \apj, 460, 136

\bibitem[{{Nagai} \& {Miyamoto}(1976)}]{Nagai1976}
{Nagai}, R. \& {Miyamoto}, M. 1976, \pasj, 28, 1

\bibitem[{{Neureiter} {et~al.}(2021){Neureiter}, {Thomas}, {Saglia}, {Bender},
  {Finozzi}, {Krukau}, {Naab}, {Rantala}, \& {Frigo}}]{Neureiter2021}
{Neureiter}, B., {Thomas}, J., {Saglia}, R., {et~al.} 2021, \mnras, 500, 1437

\bibitem[{{Pilawa} {et~al.}(2022){Pilawa}, {Liepold}, {Delgado Andrade},
  {Walsh}, {Ma}, {Quenneville}, {Greene}, \& {Blakeslee}}]{Pilawa2022}
{Pilawa}, J.~D., {Liepold}, C.~M., {Delgado Andrade}, S.~C., {et~al.} 2022,
  \apj, 928, 178

\bibitem[{Poci {et~al.}(2016)Poci, Cappellari, \& McDermid}]{Poci2016}
Poci, A., Cappellari, M., \& McDermid, R.~M. 2016, MNRAS, 467:, 1397,2017

\bibitem[{{Poci} {et~al.}(2021){Poci}, {McDermid}, {Lyubenova}, {Zhu}, {van de
  Ven}, {Iodice}, {Coccato}, {Pinna}, {Corsini}, {Falc{\'o}n-Barroso},
  {Gadotti}, {Grand}, {Fahrion}, {Mart{\'\i}n-Navarro}, {Sarzi}, {Viaene}, \&
  {de Zeeuw}}]{Poci2021}
{Poci}, A., {McDermid}, R.~M., {Lyubenova}, M., {et~al.} 2021, \aap, 647, A145

\bibitem[{{Poci} {et~al.}(2019){Poci}, {McDermid}, {Zhu}, \& {van de
  Ven}}]{Poci2019}
{Poci}, A., {McDermid}, R.~M., {Zhu}, L., \& {van de Ven}, G. 2019, \mnras,
  487, 3776

\bibitem[{{Poci} \& {Smith}(2022)}]{Poci2022}
{Poci}, A. \& {Smith}, R.~J. 2022, \mnras, 512, 5298

\bibitem[{{Qian} {et~al.}(1995){Qian}, {de Zeeuw}, {van der Marel}, \&
  {Hunter}}]{Qian1995}
{Qian}, E.~E., {de Zeeuw}, P.~T., {van der Marel}, R.~P., \& {Hunter}, C. 1995,
  \mnras, 274, 602

\bibitem[{{Quenneville} {et~al.}(2022){Quenneville}, {Liepold}, \&
  {Ma}}]{quenneville2022}
{Quenneville}, M.~E., {Liepold}, C.~M., \& {Ma}, C.-P. 2022, \apj, 926, 30

\bibitem[{{Richstone} \& {Tremaine}(1984)}]{Richstone1984}
{Richstone}, D.~O. \& {Tremaine}, S. 1984, \apj, 286, 27

\bibitem[{{Rix} {et~al.}(1997){Rix}, {de Zeeuw}, {Cretton}, {van der Marel}, \&
  {Carollo}}]{Rix1997}
{Rix}, H.-W., {de Zeeuw}, P.~T., {Cretton}, N., {van der Marel}, R.~P., \&
  {Carollo}, C.~M. 1997, \apj, 488, 702

\bibitem[{{Roberts} {et~al.}(2021){Roberts}, {Bentz}, {Vasiliev}, {Valluri}, \&
  {Onken}}]{Roberts2021}
{Roberts}, C.~A., {Bentz}, M.~C., {Vasiliev}, E., {Valluri}, M., \& {Onken},
  C.~A. 2021, \apj, 916, 25

\bibitem[{{Rusli} {et~al.}(2013){Rusli}, {Thomas}, {Saglia}, {Fabricius},
  {Erwin}, {Bender}, {Nowak}, {Lee}, {Riffeser}, \& {Sharp}}]{Rusli2013}
{Rusli}, S.~P., {Thomas}, J., {Saglia}, R.~P., {et~al.} 2013, \aj, 146, 45

\bibitem[{{Rybicki}(1987)}]{Rybicki1987}
{Rybicki}, G.~B. 1987, in IAU Symposium, Vol. 127, Structure and Dynamics of
  Elliptical Galaxies, ed. P.~T. {de Zeeuw}, 397

\bibitem[{{Santucci} {et~al.}(2022){Santucci}, {Brough}, {van de Sande},
  {McDermid}, {van de Ven}, {Zhu}, {D'Eugenio}, {Bland-Hawthorn}, {Barsanti},
  {Bryant}, {Croom}, {Davies}, {Green}, {Lawrence}, {Lorente}, {Owers}, {Poci},
  {Richards}, {Thater}, \& {Yi}}]{Santucci2022}
{Santucci}, G., {Brough}, S., {van de Sande}, J., {et~al.} 2022, arXiv
  e-prints, arXiv:2203.03648

\bibitem[{{Sarzi} {et~al.}(2018){Sarzi}, {Spiniello}, {La Barbera},
  {Krajnovi{\'c}}, \& {van den Bosch}}]{Sarzi2018}
{Sarzi}, M., {Spiniello}, C., {La Barbera}, F., {Krajnovi{\'c}}, D., \& {van
  den Bosch}, R. 2018, \mnras, 478, 4084

\bibitem[{{Satoh}(1980)}]{Satoh1980}
{Satoh}, C. 1980, \pasj, 32, 41

\bibitem[{{Schwarzschild}(1979)}]{Schwarzschild1979}
{Schwarzschild}, M. 1979, \apj, 232, 236

\bibitem[{{Scott} {et~al.}(2013){Scott}, {Cappellari}, {Davies}, {Kleijn},
  {Bois}, {Alatalo}, {Blitz}, {Bournaud}, {Bureau}, {Crocker}, {Davis}, {de
  Zeeuw}, {Duc}, {Emsellem}, {Khochfar}, {Krajnovi{\'c}}, {Kuntschner},
  {McDermid}, {Morganti}, {Naab}, {Oosterloo}, {Sarzi}, {Serra}, {Weijmans}, \&
  {Young}}]{Scott2013}
{Scott}, N., {Cappellari}, M., {Davies}, R.~L., {et~al.} 2013, \mnras, 432,
  1894

\bibitem[{{Syer} \& {Tremaine}(1996)}]{Syer1996}
{Syer}, D. \& {Tremaine}, S. 1996, \mnras, 282, 223

\bibitem[{{Thater} {et~al.}(2017){Thater}, {Krajnovi{\'c}}, {Bourne},
  {Cappellari}, {de Zeeuw}, {Emsellem}, {Magorrian}, {McDermid}, {Sarzi}, \&
  {van de Ven}}]{Thater2017}
{Thater}, S., {Krajnovi{\'c}}, D., {Bourne}, M.~A., {et~al.} 2017, \aap, 597,
  A18

\bibitem[{{Thater} {et~al.}(2019){Thater}, {Krajnovi{\'c}}, {Cappellari},
  {Davis}, {de Zeeuw}, {McDermid}, \& {Sarzi}}]{Thater2019}
{Thater}, S., {Krajnovi{\'c}}, D., {Cappellari}, M., {et~al.} 2019, \aap, 625,
  A62

\bibitem[{{Thater} {et~al.}(2022){Thater}, {Krajnovi{\'c}}, {Weilbacher},
  {Nguyen}, {Bureau}, {Cappellari}, {Davis}, {Iguchi}, {McDermid}, {Onishi},
  {Sarzi}, \& {van de Ven}}]{Thater2022}
{Thater}, S., {Krajnovi{\'c}}, D., {Weilbacher}, P.~M., {et~al.} 2022, \mnras,
  509, 5416

\bibitem[{{Thomas} {et~al.}(2007){Thomas}, {Saglia}, {Bender}, {Thomas},
  {Gebhardt}, {Magorrian}, {Corsini}, \& {Wegner}}]{Thomas2007}
{Thomas}, J., {Saglia}, R.~P., {Bender}, R., {et~al.} 2007, \mnras, 382, 657

\bibitem[{{Valluri} {et~al.}(2004){Valluri}, {Merritt}, \&
  {Emsellem}}]{Valluri2004}
{Valluri}, M., {Merritt}, D., \& {Emsellem}, E. 2004, \apj, 602, 66

\bibitem[{van~de Ven {et~al.}(2008)van~de Ven, de~Zeeuw, \& van~den
  Bosch}]{Ven2008}
van~de Ven, G., de~Zeeuw, P.~T., \& van~den Bosch, R. C.~E. 2008, \mnras, 385,
  614

\bibitem[{van~den Bosch(2016)}]{Bosch2016}
van~den Bosch, R. C.~E. 2016, The Astrophysical Journal, 831, 134

\bibitem[{{van den Bosch} \& {de Zeeuw}(2010)}]{vandenBosch2010}
{van den Bosch}, R.~C.~E. \& {de Zeeuw}, P.~T. 2010, \mnras, 401, 1770

\bibitem[{{van den Bosch} {et~al.}(2008){van den Bosch}, {van de Ven},
  {Verolme}, {Cappellari}, \& {de Zeeuw}}]{vandenBosch2008}
{van den Bosch}, R.~C.~E., {van de Ven}, G., {Verolme}, E.~K., {Cappellari},
  M., \& {de Zeeuw}, P.~T. 2008, \mnras, 385, 647

\bibitem[{{van der Marel} {et~al.}(1998){van der Marel}, {Cretton}, {de Zeeuw},
  \& {Rix}}]{vanderMarel1998}
{van der Marel}, R.~P., {Cretton}, N., {de Zeeuw}, P.~T., \& {Rix}, H.-W. 1998,
  \apj, 493, 613

\bibitem[{{Vasiliev} \& {Valluri}(2020)}]{Vasiliev2020}
{Vasiliev}, E. \& {Valluri}, M. 2020, \apj, 889, 39

\bibitem[{{Verolme} {et~al.}(2002){Verolme}, {Cappellari}, {Copin}, {van der
  Marel}, {Bacon}, {Bureau}, {Davies}, {Miller}, \& {de Zeeuw}}]{Verolme2002}
{Verolme}, E.~K., {Cappellari}, M., {Copin}, Y., {et~al.} 2002, \mnras, 335,
  517

\bibitem[{{Walsh} {et~al.}(2012){Walsh}, {van den Bosch}, {Barth}, \&
  {Sarzi}}]{Walsh2012}
{Walsh}, J.~L., {van den Bosch}, R.~C.~E., {Barth}, A.~J., \& {Sarzi}, M. 2012,
  \apj, 753, 79

\bibitem[{{Zhu} {et~al.}(2020){Zhu}, {van de Ven}, {Leaman}, {Grand},
  {Falc{\'o}n-Barroso}, {Jethwa}, {Watkins}, {Mao}, {Poci}, {McDermid}, \&
  {Nelson}}]{Zhu2020}
{Zhu}, L., {van de Ven}, G., {Leaman}, R., {et~al.} 2020, \mnras, 496, 1579

\bibitem[{{Zhu} {et~al.}(2022){Zhu}, {van de Ven}, {Leaman}, {Pillepich},
  {Coccato}, {Ding}, {Falc{\'o}n-Barroso}, {Iodice}, {Navarro}, {Pinna},
  {Corsini}, {Gadotti}, {Fahrion}, {Lyubenova}, {Mao}, {McDermid}, {Poci},
  {Sarzi}, \& {de Zeeuw}}]{Zhu2022}
{Zhu}, L., {van de Ven}, G., {Leaman}, R., {et~al.} 2022, arXiv e-prints,
  arXiv:2203.15822

\bibitem[{{Zhu} {et~al.}(2018{\natexlab{a}}){Zhu}, {van de Ven},
  {M{\'e}ndez-Abreu}, \& {Obreja}}]{Zhu2018a}
{Zhu}, L., {van de Ven}, G., {M{\'e}ndez-Abreu}, J., \& {Obreja}, A.
  2018{\natexlab{a}}, \mnras, 479, 945

\bibitem[{{Zhu} {et~al.}(2018{\natexlab{b}}){Zhu}, {van den Bosch}, {van de
  Ven}, {Lyubenova}, {Falc{\'o}n-Barroso}, {Meidt}, {Martig}, {Shen}, {Li},
  {Yildirim}, {Walcher}, \& {Sanchez}}]{Zhu2018b}
{Zhu}, L., {van den Bosch}, R., {van de Ven}, G., {et~al.} 2018{\natexlab{b}},
  \mnras, 473, 3000

\end{thebibliography}
\bibliographystyle{aa}

\begin{appendix} 

\section{Impact of incorrect orbit mirroring for a model with close to axisymmetric potential} 

\label{apxA}

\begin{figure*}%
\centering
\includegraphics[width=0.85\textwidth]{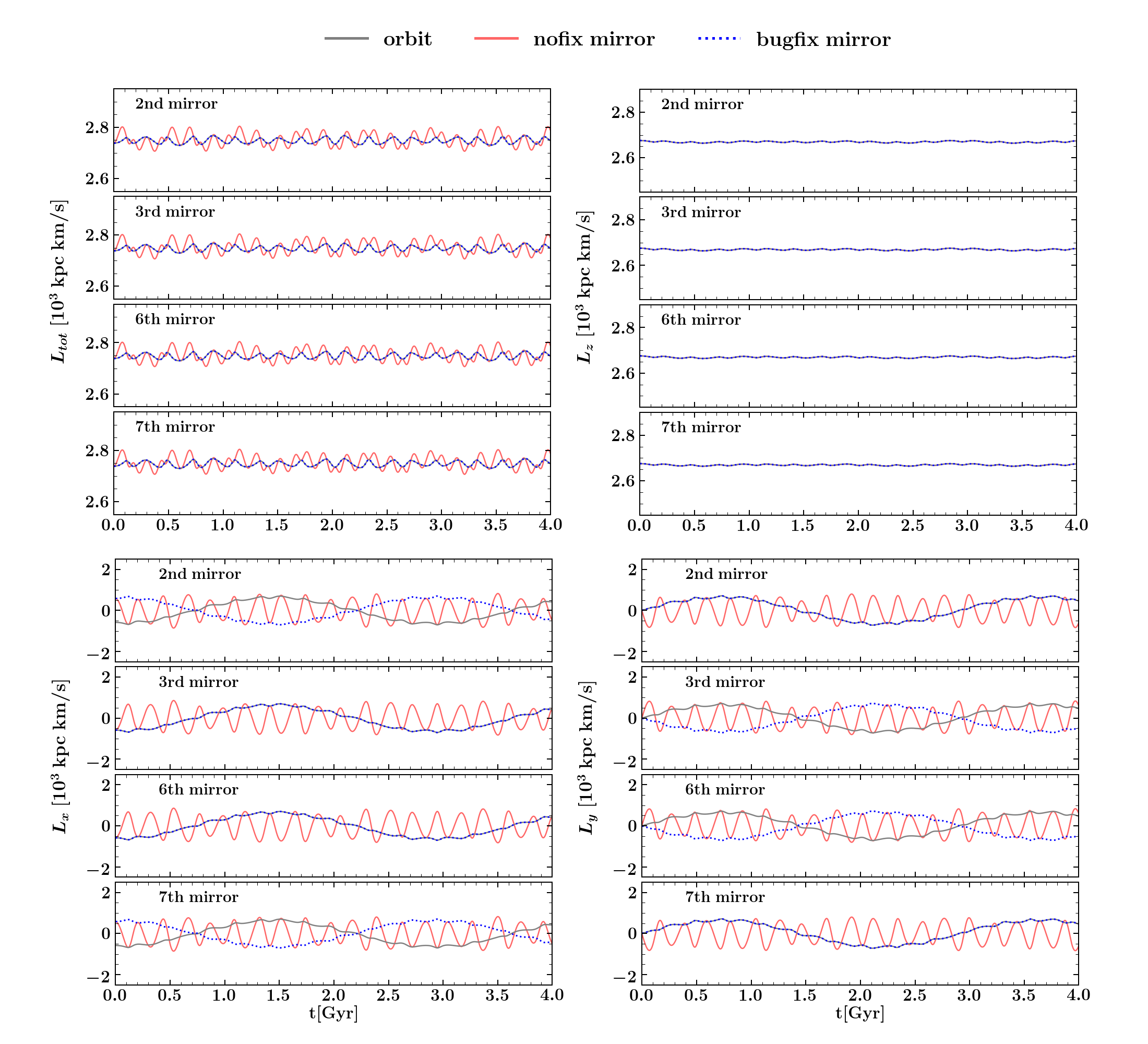} \\
\caption[]{Similar to Fig. \ref{fig:orb}  but for a short-axis tube orbit in a model with axisymmetric potential. The orbit is shown for 9 (out of 200) revolutions. The time of one orbital period is $\sim 0.44$ Gyr. }%
\label{fig:orbA}%
\end{figure*}

\begin{figure*}%
\centering
\includegraphics[width=0.85\textwidth]{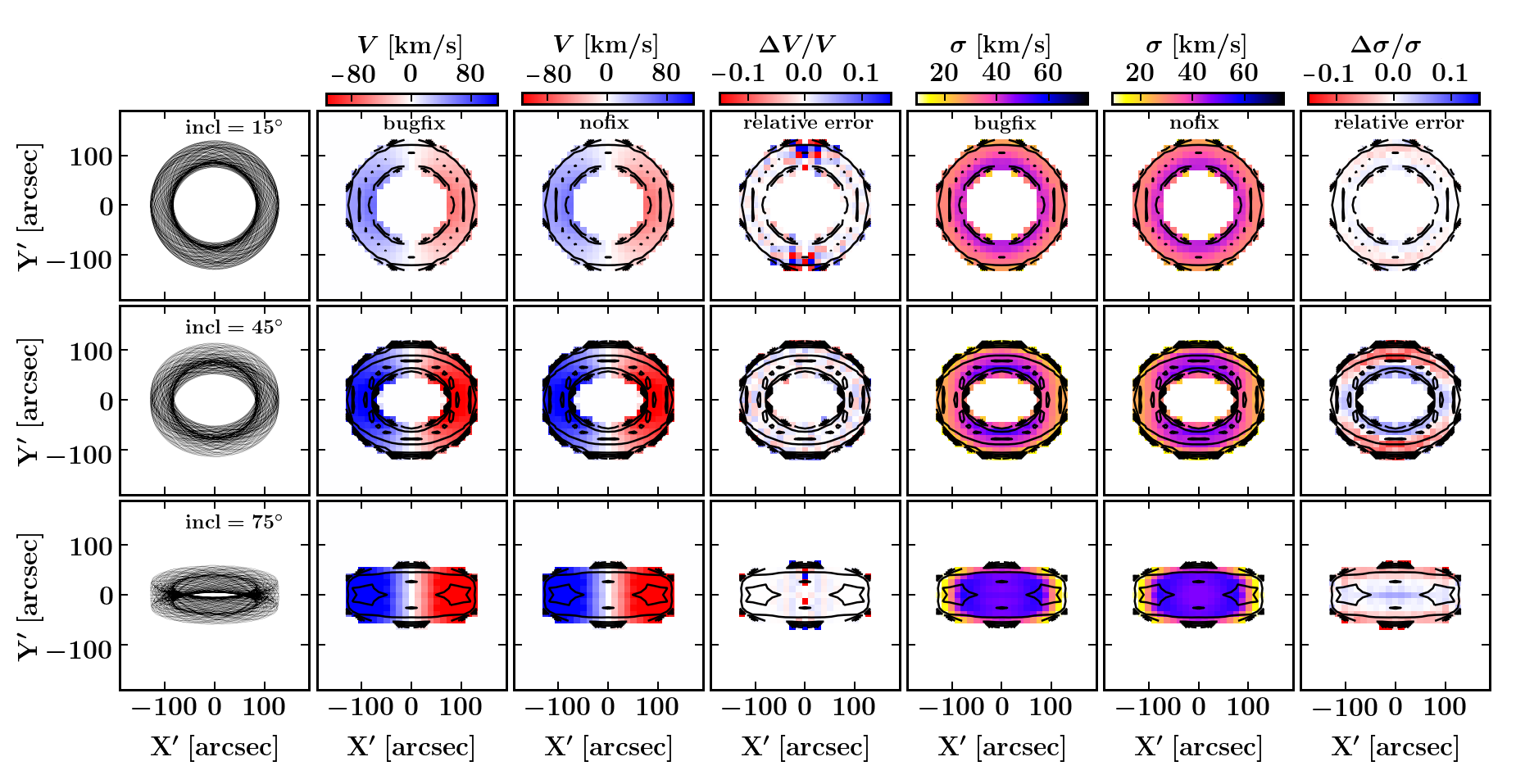} \caption[]{Similar to Fig. \ref{fig:orb2} but for a short-axis tube orbit in a model with axisymmetric potential. This orbit
 has a large weight in the model that is discussed in section 5.}%
\label{fig:orb_TA}%
\end{figure*}

We consider here the effects of the correct and incorrect mirroring scheme for a galactic potential in the axisymmetric limit, derived from the luminosity model of the simulated Auriga galaxy halo 6 \citet{Zhu2020}. Figures \ref{fig:orbA} and \ref{fig:orb_TA} show the impact of the incorrect orbit mirroring for a single orbit, in an analogous way to what is shown in Fig. \ref{fig:orb} and \ref{fig:orb2} in Section 2.

\begin{figure*}%
\centering
\includegraphics[width=0.85\textwidth]{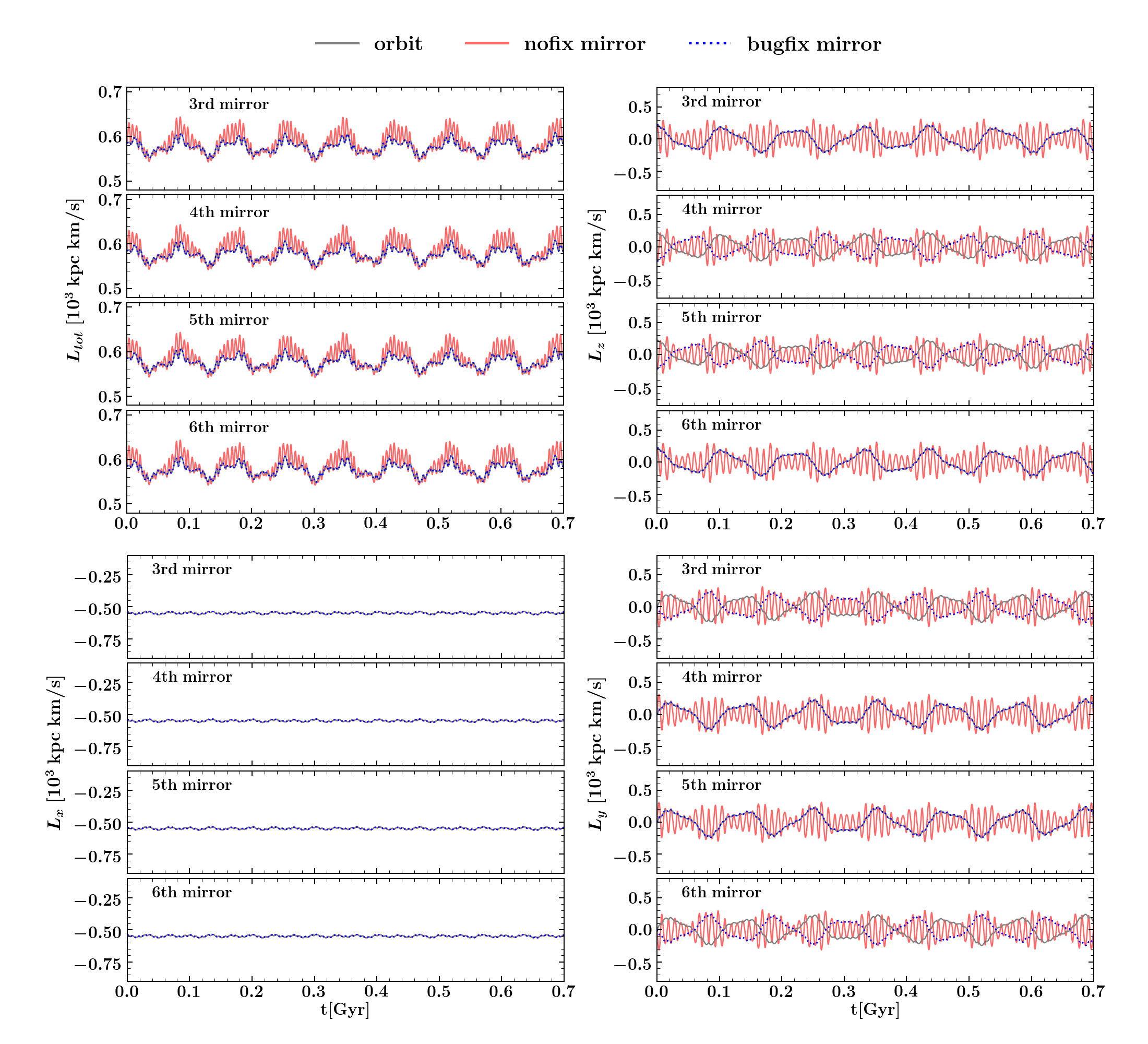} \\
\caption[]{Similar to Fig. \ref{fig:orb}  but for an outer long-axis tube orbit in the model with triaxial potential. The orbit is shown for 40 (out of 200) revolutions. The time of one orbital period is $\sim 0.018$ Gyr.}%
\label{fig:orbB}%
\end{figure*}

\begin{figure*}%
\centering
\includegraphics[width=0.85\textwidth]{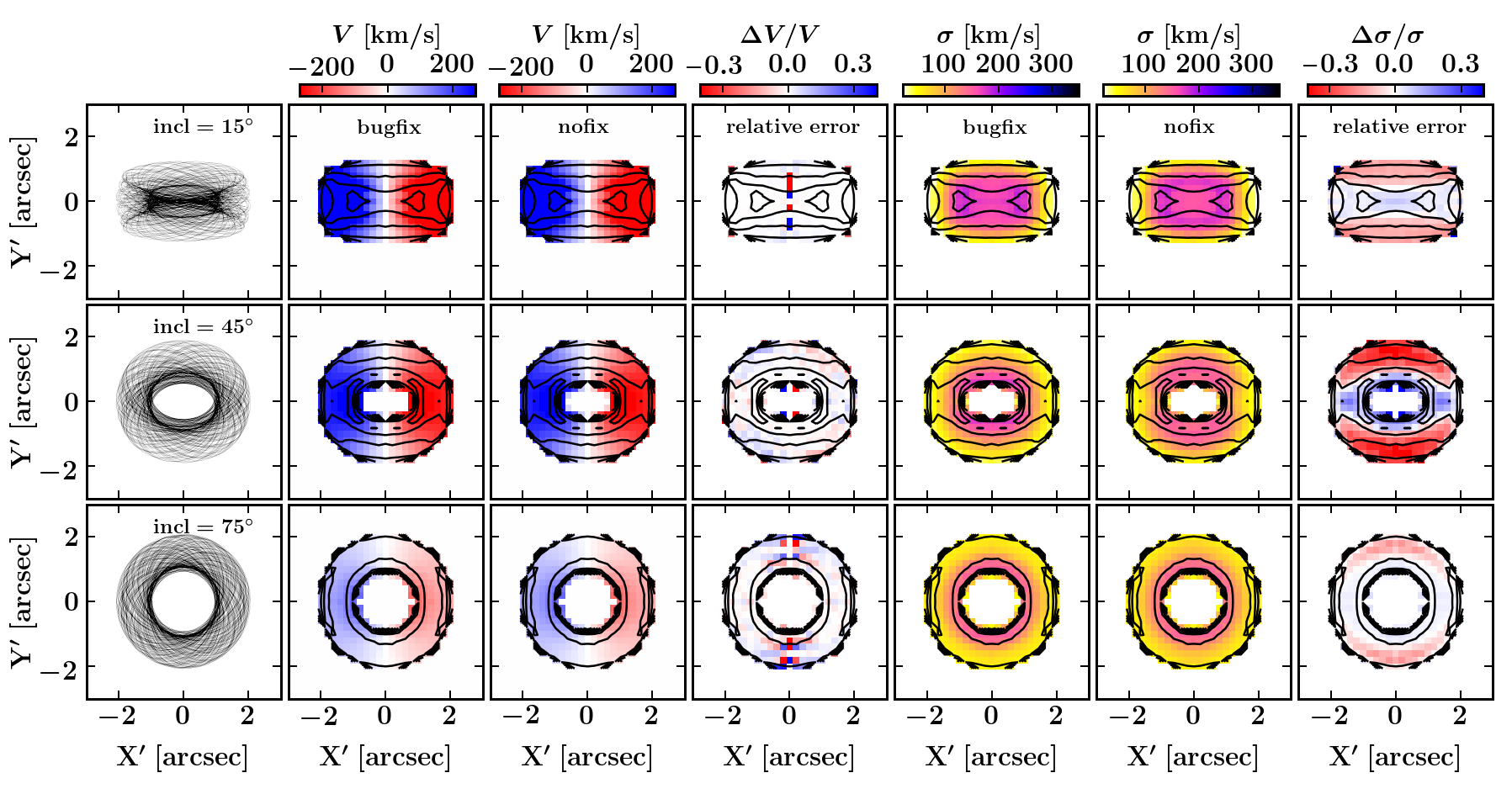}
\caption[]{Similar to Fig. \ref{fig:orb2} but for an outer long-axis tube orbit in the model with triaxial potential.}%
\label{fig:orb_TB}%
\end{figure*}

To further explore the more complex triaxial case shown in Section 2, we show in Fig. \ref{fig:orbB} and \ref{fig:orb_TB} the impact of the mirroring on a long-axis tube orbit.

\section{Orbit distribution of PGC 046832 } 

\begin{figure*}%
\centering
\includegraphics[width=9.0cm]{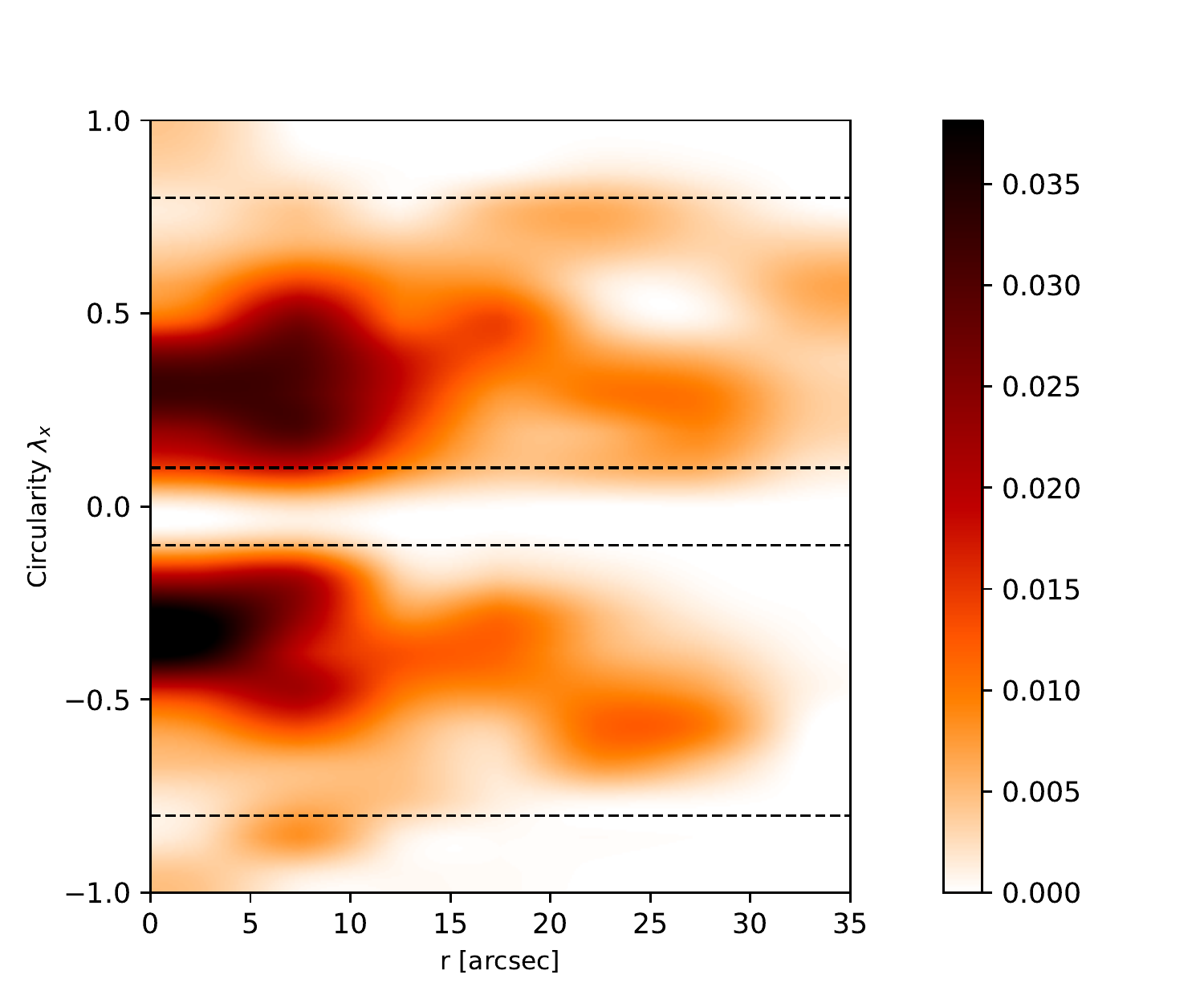}
\includegraphics[width=9.0cm]{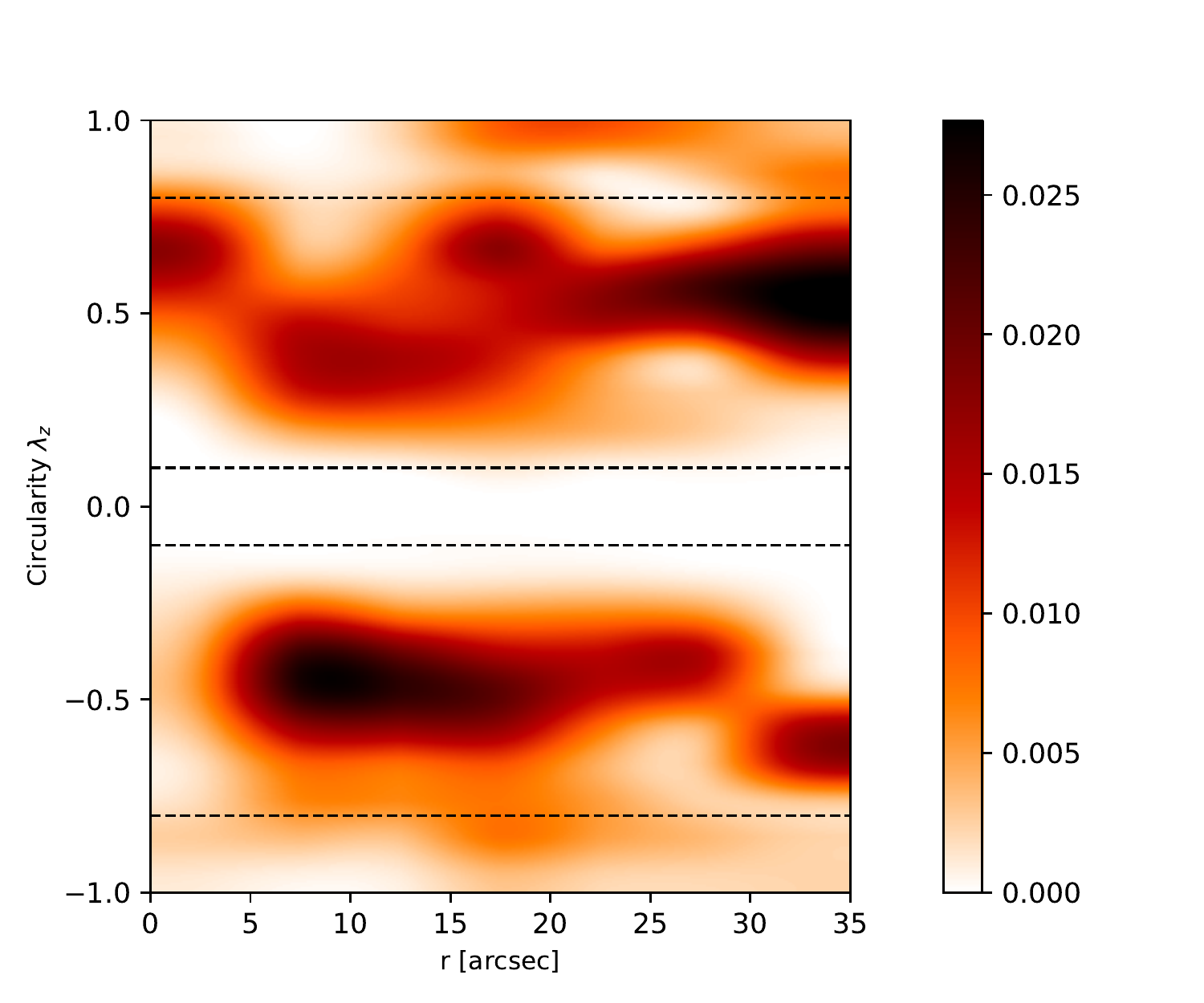}
\caption[]{Average orbital circularity of tube orbits as a function of radius for all models consistent with the best fit model of PGC 046832 (discussed in Section 4). Darker colours imply a higher density of orbits. The dashed lines separate hot orbits, warm orbits and cold orbits.} %
	\label{fig:pgc_circ}%
\end{figure*}

In Fig.~\ref{fig:pgc_circ} we show the stellar orbit distribution obtained for the best-fit model of PGC 046832, when using a Schwarzschild code with correct orbit mirroring.

\end{appendix}
\end{document}